\documentclass{article}
\usepackage{amsfonts,amssymb,amsmath}
\usepackage{color, xcolor}
\usepackage{listings}
\usepackage{graphicx}
\usepackage{hyperref}
\usepackage{url}
\usepackage{ulem}

\newcommand {\susy}{\mathfrak{susy}}

\newcommand{\Imm}{\mathrm{Im}}

\newcommand {\C}{\mathbb{C}}

\newcommand {\G}{\mathcal{G}}
\newcommand {\Spin}{\mathrm{Spin}}

\newcommand {\SO}{\mathrm{SO}}
\newcommand {\SU}{\mathrm{SU}}

\newcommand {\sso}{\mathfrak{so}}
\newcommand {\gl}{\mathfrak{gl}}

\renewcommand {\sl}{\mathfrak{sl}}
\newcommand {\s}{\mathfrak{s}}

\newcommand{\Sym}{\mathrm{Sym}}
\newcommand {\Ker}{\mathrm{Ker}}

\newcommand{\beq}{\begin{equation}}
\newcommand{\eeq}{\end{equation}}
\newcommand{\beqr}{\begin{eqnarray}}
\newcommand{\eeqr}{\end{eqnarray}}

\stepcounter{equation}

\begin{document}
\title {Homology of Lie algebra of supersymmetries and of super Poincare Lie algebra\thanks {The work was partially supported by NSF grant DMS-0805989}}
\author{M. V. Movshev\\Stony Brook University\\Stony Brook, NY 11794-3651, USA
\\ A. Schwarz\\ Department of Mathematics\\ University of
California \\ Davis, CA 95616, USA, \\Renjun Xu \\Department of Physics, University of California \\ Davis, CA 95616, USA }
\date{\today}
\maketitle

Abstract { We study the homology and cohomology groups of super Lie algebra of supersymmetries  and of super Poincare Lie algebra. We give complete answers for (non-extended) supersymmetry  in all dimensions $\leq 11$. For dimensions $D=10,11$ we describe also the cohomology of  reduction of supersymmetry Lie algebra to lower dimensions. Our  methods can be applied  to extended supersymmetry algebra.}

\section {Introduction}

In present paper we will analyze homology and cohomology groups of the super Lie algebra of supersymmetries and of super Poincare Lie algebra. We came to this problem studying supersymmetric
deformations of maximally supersymmetric gauge theories ~\cite {MS}; however, this problem arises also in different situations, in particular, in supergravity~\cite{Supergravity}.  In low dimensions it was studied in~\cite{Brandt}. The cohomology of supersymmetry Lie algebra appeared in the analysis of supersymmetric invariants in \cite {BH} (it was denoted there by the symbol $H^{p,q}_t$).  The cohomology groups calculated  below appear also in pure spinor formalism of ten-dimensional supersymmetric gauge theory  in ~\cite {Be}. Some of results of present paper were derived by more elementary methods in our previous paper \cite {MSX}.

Let us recall the definition of Lie  algebra cohomology. We start with super Lie algebra $\cal G$ with generators $e_A$ and structure constants $f_{AB}^K$. We introduce ghost variables $C^A$ with
parity opposite to the parity of generators $e_A$ and consider the algebra $E$ of polynomial functions of these variables. (In more invariant way we can say that $E$ consists of polynomial functions
on linear superspace $\Pi \cal G$ . The algebra $E$ is graded by the degree of polynomial. We define a derivation $d$ on $E$ by the formula
   $d=\frac{1}{2}f^K_{AB}C^AC^B\frac{\partial}{\partial C^K}.$

This operator is a differential (i.e. it changes the parity and obeys $d^2=0.$) We define the cohomology of $\cal G$ using this differential:
$$H^{\bullet}({\cal G})= \Ker d/\Imm d.$$
The definition of homology of $\cal G$ is dual to the definition of cohomology: instead of $E$ we consider its dual space $E^*$ that can be considered as the space of functions of dual ghost variables
$c_A$; the differential $\partial$ on $E^*$ is defined as an operator adjoint to $d$. The homology $H_{\bullet}(\cal G)$ is dual to the cohomology $H^{\bullet}(\cal G)$. We will work with cohomology, but our results can be interpreted in the language of homology.

Notice that we can multiply cohomology
classes, i.e. $H^{\bullet}(\cal G)$ is an algebra.

The group $Aut (\cal G)$ of automorphisms of $\cal G$ acts on $E$ and commutes with the differential, therefore it acts also on homology and cohomology.  We will be interested in this action. In other words we calculate cohomology as representation of this group  or as a representation of  its Lie algebra $aut (\cal G)$ (as  an $aut (\cal G)$-module).   For every graded module  $E$ we can define its Euler characteristic $\chi (E)$ as a virtual module  $\sum (-1)^k E_k$
(as an alternating sum of its graded components in the sense of K-theory). Euler characteristic of graded differential module  coincides with Euler characteristic of its homology. This allows us to calculate the Euler characteristic of Lie algebra cohomology as virtual representation  (virtual $aut (\cal G)$-module) . {\footnote {Instead of virtual modules we can talk about virtual representations of $Aut (\cal G)$ (elements of representation ring). If the group $Aut (\cal G)$ is compact the representation ring can be identified  with the ring of characters.}}  If the cohomology does not vanish only in one degree the Euler characteristic gives a complete answer for cohomology.

The super Lie algebra of supersymmetries has odd generators $e_{\alpha}$ and even generators $P_m$ ; the only non-trivial  commutation relation is
$$[e_{\alpha}, e_{\beta}]_+=\Gamma _{\alpha \beta}^m P_m.$$
The coefficients in this relation are expressed in terms of Dirac Gamma matrices (see e.g. \cite{Deligne}  for mathematical introduction). The space $E$ used in the definition of cohomology  (cochain complex) consists here of polynomial functions of even ghost variables $t^{\alpha}$ and odd ghost
variables $ c^m$; the differential has the form
\begin {equation}
\label {sud}
d=\frac{1}{2}\Gamma _{\alpha \beta}^mt^{\alpha}t^{\beta}\frac{\partial}{\partial c^m}.
\end {equation}
 The space $E$ is double-graded (one can consider the degree with respect to $t^{\alpha}$ and the degree with respect to $c^m$). In more invariant form we can say that {\footnote {We use the notation
$\Sym^m$ for symmetric tensor power and the notation $\Lambda ^n $ for exterior power}}
$$E=\bigoplus \Sym^m S\otimes \Lambda ^n V$$
where $S$ stands for spinorial  representation of orthogonal group, $V$ denotes vector representation of this group and Gamma-matrices specify an intertwiner $V\to \Sym^2 S$. The differential $d$ maps
$\Sym^m S\otimes \Lambda ^n V$ into $\Sym^{m+2} S\otimes \Lambda ^{n-1}V$. The above description  can be applied to any dimension and to any signature of the metric used in the definition of orthogonal
group, however,  spinorial representation is dimension-dependent. {\footnote {Recall that orthogonal group $\SO(2n)$ has two irreducible two-valued complex representations called semi-spin representations (left spinors and right spinors), the orthogonal group $\SO(2n+1)$ has one irreducible two-valued complex spin representation. One says that a real representation is spinorial if after extension of scalars to $\mathbb{C}$ it becomes a sum of spin or semi-spin representations. (We follow the terminology of \cite {Deligne}.)}}  The group $\SO(n)$
can be considered as a (subgroup) of the group of automorphisms of supersymmetry Lie algebra and therefore it acts on its cohomology . The action of $\SO (n)$ is two-valued, hence it would be more precise to talk about action of its two-sheeted covering $\Spin (n)$ or about action of its Lie algebra $\sso(n)$. We will work with complex representations and complex Lie algebras; this  does not change  the cohomology

We will consider also homology and cohomology of reduced Lie algebra of supersymmetries (or more precisely the Lie algebra of supersymmetries in dimension $n$ reduced to the dimension $d$). This algebra has the same odd generators $e_{\alpha}$ as  the Lie algebra of supersymmetries in dimension $n$, but only $d\leq \dim V$ even generators $P_1,...,P_d$; the commutation relations are the same as in unreduced algebra.  In this case the cohomology is a representation of $\Spin(d)\times \Spin(n-d)$.

 The double grading on $E$ induces double grading on cohomology. However, instead of the degrees $m$ and $n$ it is more convenient to use the degrees
$k=m+2n$ and $n$ because the differential preserves $k$ and therefore the problem of calculation of cohomology can be solved for every $k$ separately. It important to notice that the differential
commutes with multiplication by a polynomial in $t^{\alpha}$, therefore the cohomology is a module over the polynomial ring ${\C}[t^1,..., t^{\alpha},...].$ (Moreover, it is an algebra
over this ring.) The cohomology is infinite-dimensional as a vector space, but it has a finite number of generators as a ${\C}[t^1,..., t^{\alpha},...]$-module (this follows from the fact that the
polynomial ring is noetherian). One of the  problems we would like to solve is the description of these generators. If cohomology classes of cocycles $z_1,...,z_N$ generate the cohomology then every cohomology class can be represented by a cocycle of the form $p_1z_1+...+p_Nz_N$ where $p_1,..., p_N$ belong to  ${\C}[t^1,..., t^{\alpha},...]$.

Notice that the cohomology of Lie algebra of supersymmetries can be interpreted as homology of Koszul complex~{\footnote {To define the homology of Koszul complex corresponding to a sequence of functions $f^1(t),...,f^n(t)$ one considers the differential $d=f^m(t)\frac{\partial}{\partial \xi _m}$ where $\xi _1,...,\xi _n$ are odd variables.}} corresponding to to a sequence of functions $f^m(t)=\frac{1}{2}\Gamma _{\alpha \beta}^mt^{\alpha}t^{\beta}$. This allows us to use software \cite {mac2} to calculate the dimensions of cohomology groups. However, we are interested in more complicated problem- in the description of decomposition of cohomology groups in direct sum of irreducible representations of the group of automorphisms $Aut$ or its Lie algebra $aut$.
We use \cite{LiEcode} for such calculations.

The paper is organized as follows. We start with the description of cohomology of Lie algebra of supersymmetries in dimension $10$ (Sec.\ref{Sec:10D}) and in dimension $11$ (Sec.\ref{Sec:11D}). In the next sections we describe cohomology of dimensional reductions of ten-dimensional algebra of supersymmetries (Sec.\ref{Sec:redux10D}) and of eleven-dimensional supersymmetries (Sec.\ref{Sec:redux11D}). Section \ref{Sec:otherD} contains the results about Lie algebras of supersymmetries in dimensions $\leq 9$. Section \ref{Sec:Calculations} is devoted to the explanation of methods we are using.  Section \ref{Sec:superPoincare} is devoted to cohomology of super Poincare Lie algebra.  The  paper contains two appendices that will be omitted in printed version.  In Appendix A we describe the decomposition of
free resolution in direct sum of representations of the automorphism group. Appendix B gives more detail about our calculations.

{\bf Acknowledgements}

We are indebted to F. Brandt for useful correspondence.

\section {D=10\label{Sec:10D}}
We will  start with ten-dimensional case; in this case the  spinorial representation in the definition of Lie algebra of supersymmetries
should be considered as one of  two irreducible
 16-dimensional representations of $\Spin(10)$ (in Minkowski space  the spinors are Majorana-Weyl spinors).  The Lie algebra of automorphisms $aut$ is $\sso (10)$.

We will describe representations of the Lie algebra $\sso{(10)}$ in  the cohomology of the Lie algebra of  supersymmetries in ten dimensions. As usual the representations are labeled by coordinates of their
highest weight (see e.g. \cite{FultonRep} for details). The vector representation $V$ has the highest weight $[1,0,0,0,0]$, the irreducible spinor representations have highest weights $[0,0,0,0,1]$,$[0,0,0,1,0]$; we assume that the highest weight of $S$ is $[0,0,0,0,1]$. The description
of graded component of cohomology group with gradings $k=m+2n$ and $n$ is given by the formulas for $H^{k,n}$ (for $n\geq 6$, $H^{k,n}$ vanishes)
\beqr
H^{k,0}&=& [0,0,0,0,k] \label{Hk,0}\\
H^{k,1}&=& [0,0,0,1,k-3] \\
H^{k,2}&=& [0,0,1,0,k-6] \\
H^{k,3}&=& [0,1,0,0,k-8] \\
H^{k,4}&=& [1,0,0,0,k-10] \\
H^{k,5}&=& [0,0,0,0,k-12] \label{Hk,5}
\eeqr
The only special case is when $k=4$, there is one additional term, a scalar, for $H^{4,1}$.
\beq
H^{4,1}=[0,0,0,0,0]\oplus[0,0,0,1,1] \label{H4,1}
\eeq
The $\SO(10)$-invariant part is in $H^{0,0}$, $H^{12,5}$, and $H^{4,1}$.

The dimensions of these cohomology groups are encoded in series $P_n(\tau)=\sum_k \dim H^{k,n}\tau ^k$ (Poincare series) that can be calculated by means of \cite {mac2}:
\beqr
P_0(\tau)&=&\frac{{\tau}^3 + 5{\tau}^2 + 5{\tau} + 1}{(1 - {\tau})^{11}}, \\
P_1(\tau)&=&(16 {\tau}^3 + 35 {\tau}^4 - {\tau}^5 + 55 {\tau}^6 - 165 {\tau}^7 + 330 {\tau}^8 - 462 {\tau}^9 + 462 {\tau}^{10}\nonumber\\
        && - 330 {\tau}^{11} + 165 {\tau}^{12} - 55 {\tau}^{13} + 11 {\tau}^{14} - {\tau}^{15})/(1 - {\tau})^{11},\\
P_2(\tau)&=&(120 {\tau}^6 - 120 {\tau}^7 + 330 {\tau}^8 - 462 {\tau}^9 + 462 {\tau}^{10} - 330 {\tau}^{11} \nonumber\\
        &&+ 165 {\tau}^{12} - 55 {\tau}^{13} + 11 {\tau}^{14} - {\tau}^{15})/(1 - {\tau})^{11},\\
P_3(\tau)&=&\frac{45 {\tau}^8 + 65 {\tau}^9 + 11 {\tau}^{10} - {\tau}^{11}}{(1 - {\tau})^{11}},\\
P_4(\tau)&=&\frac{10 {\tau}^{10} + 34 {\tau}^{11} + 16 {\tau}^{12}}{(1 - {\tau})^{11}},\\
P_5(\tau)&=&\frac{{\tau}^{12} + 5 {\tau}^{13} + 5 {\tau}^{14} + {\tau}^{15}}{(1 - {\tau})^{11}}
\eeqr

The cohomology regarded as ${\C}[t^1,..., t^{\alpha},...]$-module is generated by the scalar considered as an element of $H^{0,0}$ and by
\beqr
 \left[t^{\alpha}c_m \Gamma^m_{\alpha\beta}\right]\in H^{3,1}, \nonumber\\
 \left[t^{\alpha}t^{\beta}c_m c_n \Gamma^{mnklr}_{\alpha\beta}\right]\in H^{6,2}, \nonumber \\
 \left[t^{\alpha}t^{\beta}c_m c_n c_k \Gamma^{mnklr}_{\alpha\beta}\right]\in H^{8,3}, \nonumber \\
 \left[t^{\alpha}t^{\beta}c_m c_n c_k c_l \Gamma^{mnklr}_{\alpha\beta}\right]\in H^{10,4}, \nonumber \\
 \left[t^{\alpha}t^{\beta}c_m c_n c_k c_l c_r \Gamma^{mnklr}_{\alpha\beta}\right]\in H^{12,5}. \nonumber
\eeqr
Here  $[a]$ denotes the cohomological class of cocycle $a$.

The GAMMA package~\cite{GAMMA}  was used to verify that the expression above are cocycles .
\section {D=11\label{Sec:11D}}
Now we consider the eleven-dimensional case; in this case  the  spinorial representation in the definition of supersymmetry Lie algebra
should be considered as one irreducible 32-dimensional spinor representations of $\Spin(11)$ (Dirac spinors).  As usual we work with complex representations and complex Lie algebras.
.

We will describe representations of $Aut (\G)=\sso(11)$ in the  the cohomology of the Lie algebra of  supersymmetries. As usual the representations are labeled by their
highest weight. The vector representation $V$ has the highest weight $[1,0,0,0,0]$, the irreducible spinor representations have highest weights $[0,0,0,0,1]$. The description
of graded component of cohomology group with gradings $k=m+2n$ and $n$ is given by the formulas for $H^{k,n}$ (for $n\geq 3$, $H^{k,n}$ vanishes)
\beqr
H^{k,0}&=& \overset{\left[k/2\right]}{\underset{i=0}{\oplus}}[0,i,0,0,k-2i] \label{Hk11,0}\\
H^{k,1}&=& \overset{\left[(k-4)/2\right]}{\underset{i=0}{\oplus}}[1,i,0,0,k-4-2i] \\
H^{k,2}&=& \overset{\left[(k-6)/2\right]}{\underset{i=0}{\oplus}}[0,i,0,0,k-6-2i]
\eeqr
The $\SO(11)$-invariant part is in $H^{0,0}$ and $H^{6,2}$.

The dimensions of these cohomology groups are encoded in Poincare series:
\beqr
P_0(\tau)&=&A(\tau) \\
P_1(\tau)&=&\frac{{\tau}^4(11 + 67{\tau}  + 142{\tau}^2  + 142{\tau}^3  + 67{\tau}^4  + 11{\tau}^5)}{(1 - {\tau})^{23}}, \\
P_2(\tau)&=&A(\tau){\tau}^{6}
\eeqr
where
\beq
A(\tau)=\frac{1 + 9{\tau} + 34{\tau}^2  + 66{\tau}^3  + 66{\tau}^4  + 34{\tau}^5  + 9{\tau}^6  + {\tau}^7}{(1-\tau)^{23}} \label{P11D:0}
\eeq
The cohomology regarded as ${\C}[t^1,..., t^{\alpha},...]$-module is generated by the scalar considered as an element of $H^{0,0}$ and
\beqr
 \left[t^{\alpha}t^{\beta}c_m \Gamma^{mn}_{\alpha\beta}\right]\in H^{4,1}, \nonumber\\
 \left[t^{\alpha}t^{\beta}c_m c_n \Gamma^{mn}_{\alpha\beta}\right]\in H^{6,2}, \nonumber
\eeqr

\section {Dimensional reduction from $D=10$\label{Sec:redux10D}}
Let us consider dimensional reductions of ten-dimensional Lie algebra of supersymmetries. The reduction of $\susy _{10}$ to $r$ dimensions has $16$ odd generators (supersymmetries) and $r$ even generators (here $0\leq r\leq 10$). Corresponding differential has the form (\ref {sud}) where  $\Gamma _{\alpha \beta}^m$ are ten-dimensional Dirac matrices, Greek indices take $16$ values as in unreduced case, but Roman indices take only $d$ values. The differential commutes with (two-valued)  action of the group $\SO(r)\times \SO (10-r)$, therefore this group acts on cohomology. The cohomology can be regarded as a module over $\mathbb{C}[t^1,...,t^{16}].$
Again cohomology is double graded; we use notation $H^{k,n}$ for the component having degree $m=k-2n$ with respect to $t$ and the degree $n$ with respect to $c$. The symbol $P_n(\tau)$ stands for the generating  function $P_n(\tau)=\sum_k \dim H^{k,n}\tau ^k$ (for Poincare series).
We calculate the cohomology as a representation of Lie algebra $\sso(r)\times \sso(10-r)$ and describe elements that generate it as a $\mathbb{C}[t^1,...,t^{16}]-$module. (We characterize the representation writing  Dynkin labels of the first factor, then Dynkin labels of second factor.)

\begin{itemize}
\item $r=9$,
\beqr
H^{k,0}&=& [0,0,0,k], k\neq 2\\
H^{k,1}&=& [0,0,1,k-4] \\
H^{k,2}&=& [0,1,0,k-6] \\
H^{k,3}&=& [1,0,0,k-8] \\
H^{k,4}&=& [0,0,0,k-10]
\eeqr
when $k=2$,
\beq
H^{2,0}= [0,0,0,0]\oplus[0,0,0,2]
\eeq
Groups $H^{k,n}$ with $n\geq 5$ vanish.
The $\SO(9)$-invariant part is in $H^{0,0}$, $H^{10,4}$, and $H^{2,0}$.

Generators
\beqr
 \left[t^{\alpha}t^{\beta}c_m \Gamma^{mnkl}_{\alpha\beta}\right]\in H^{4,1}, \nonumber\\
 \left[t^{\alpha}t^{\beta}c_m c_n \Gamma^{mnkl}_{\alpha\beta}\right]\in H^{6,2}, \nonumber \\
 \left[t^{\alpha}t^{\beta}c_m c_n c_k \Gamma^{mnkl}_{\alpha\beta}\right]\in H^{8,3}, \nonumber \\
 \left[t^{\alpha}t^{\beta}c_m c_n c_k c_l \Gamma^{mnkl}_{\alpha\beta}\right]\in H^{10,4}. \nonumber
\eeqr

Poincare series
\beqr
P_0(\tau)&=&({\tau}^{13} - 11{\tau}^{12} + 55{\tau}^{11} - 165{\tau}^{10} + 330{\tau}^9 - 462{\tau}^8 + 462{\tau}^7 \nonumber\\
        &&\!\!\!\!\!\!\!\!\!\!\! -330{\tau}^6 + 165{\tau}^5 - 55{\tau}^4 + 10{\tau}^3 - 6{\tau}^2 - 5{\tau} - 1)/(-1 + {\tau})^{11},\\
P_1(\tau)&=&(84 {\tau}^4 - 156 {\tau}^5 + 330 {\tau}^6 - 462 {\tau}^7 + 462 {\tau}^8 - 330 {\tau}^9  \nonumber\\
        &&{} + 165 {\tau}^{10} - 55 {\tau}^{11} + 11 {\tau}^{12} - {\tau}^{13})/(1 - {\tau})^{11},\\
P_2(\tau)&=&\frac{36 {\tau}^6 + 36 {\tau}^7}{(1 - {\tau})^{11}},\\
P_3(\tau)&=&\frac{9 {\tau}^8 + 29 {\tau}^9 + 11 {\tau}^{10} - {\tau}^{11}}{(1 - {\tau})^{11}},\\
P_4(\tau)&=&\frac{{\tau}^{10} + 5 {\tau}^{11} + 5 {\tau}^{12} + {\tau}^{13}}{(1 - {\tau})^{11}}
\eeqr

\item $r=8, k>0$,
\beqr
H^{k,0}&=& \overset{k-1}{\underset{i=1}{\oplus}}[0,0,k-i,i,k-2i] \overset{\left[k/2\right]}{\underset{i=0}{\oplus}}[0,0,k-2i,0,k]\nonumber\\
 && \overset{\left[k/2\right]}{\underset{i=0}{\oplus}}[0,0,0,k-2i,-k],\\
H^{k,1}&=& \overset{k-4}{\underset{i=0}{\oplus}}[0,1,k-4-i,i,k-4-2i], \\
H^{k,2}&=& \overset{k-6}{\underset{i=0}{\oplus}}[1,0,k-6-i,i,k-6-2i], \\
H^{k,3}&=& \overset{k-8}{\underset{i=0}{\oplus}}[0,0,k-8-i,i,k-8-2i]
\eeqr
Groups $H^{k,n}$ with $n\geq 4$ vanish.
The $\SO(8)\times \SO(2)$-invariant part is in $H^{0,0}$, and $H^{8,3}$.

Generators:
\beqr
 \left[t^{\alpha}t^{\beta}c_m \Gamma^{mnk}_{\alpha\beta}\right]\in H^{4,1}, \nonumber\\
 \left[t^{\alpha}t^{\beta}c_m c_n \Gamma^{mnk}_{\alpha\beta}\right]\in H^{6,2}, \nonumber \\
 \left[t^{\alpha}t^{\beta}c_m c_n c_k \Gamma^{mnk}_{\alpha\beta}\right]\in H^{8,3}, \nonumber
\eeqr

Poincare series
\beqr
P_0(\tau)&=&\frac{2{\tau}^5 - 6{\tau}^4 + 5{\tau}^3 - 7{\tau}^2 - 5{\tau} - 1}{(-1 + {\tau})^{11}},\\
P_1(\tau)&=&\frac{28 {\tau}^4 + 12 {\tau}^5 - 6 {\tau}^6 + 2 {\tau}^7}{(1 - {\tau})^{11}},\\
P_2(\tau)&=&\frac{8 {\tau}^6 + 24 {\tau}^7 + 6 {\tau}^8 - 2 {\tau}^9}{(1 - {\tau})^{11}},\\
P_3(\tau)&=&\frac{{\tau}^8 + 5 {\tau}^9 + 5 {\tau}^{10} + {\tau}^{11}}{(1 - {\tau})^{11}}
\eeqr

\item $r=7$,
\beqr
H^{k,0}&=& \overset{\left[k/2\right]}{\underset{i=1}{\oplus}}[0,i,k-2i,k-2i] \overset{\left[k/2\right]}{\underset{i=0}{\oplus}}[0,0,k-2i,k]\\
H^{k,1}&=& \overset{\left[(k-4)/2\right]}{\underset{i=0}{\oplus}}[1,i,k-4-2i,k-4-2i] \\
H^{k,2}&=& \overset{\left[(k-6)/2\right]}{\underset{i=0}{\oplus}}[0,i,k-6-2i,k-6-2i]
\eeqr
Groups $H^{k,n}$ with $n\geq 3$ vanish.
The $\SO(7)\times \SO(3)$-invariant part is in $H^{0,0}$, and $H^{6,2}$.

Generators:
\beqr
 \left[t^{\alpha}t^{\beta}c_m \Gamma^{mn}_{\alpha\beta}\right]\in H^{4,1}, \nonumber\\
 \left[t^{\alpha}t^{\beta}c_m c_n \Gamma^{mn}_{\alpha\beta}\right]\in H^{6,2}, \nonumber
\eeqr

Poincare series
\beqr
P_0(\tau)&=&\frac{5{\tau}^5 - 7{\tau}^4 + 8{\tau}^2 + 5{\tau} + 1}{(-1 + {\tau})^{11}},\\
P_1(\tau)&=&\frac{7 {\tau}^4 + 19 {\tau}^5 + {\tau}^6 - 3 {\tau}^7}{(1 - {\tau})^{11}},\\
P_2(\tau)&=&\frac{{\tau}^6 + 5 {\tau}^7 + 5 {\tau}^8 + {\tau}^9}{(1 - {\tau})^{11}}
\eeqr

\item $r=6$,
\beqr
H^{k,0}&=& \overset{\left[k/2\right]}{\underset{i=0}{\oplus}}\overset{k-2i}{\underset{j=0}{\oplus}}[j,i,k-j-2i,j,k-j-2i] \nonumber\\ && \overset{k-1}{\underset{i=1}{\oplus}}\overset{i-1}{\underset{j=\max\{0,2i-k\}}{\oplus}}[j,0,k-2i+j,i,k-i]\\
H^{k,1}&=& \overset{\left[(k-4)/2\right]}{\underset{i=0}{\oplus}}\overset{k-4-2i}{\underset{j=0}{\oplus}}[j,i,k-4-j-2i,j,k-4-j-2i]
\eeqr
Groups $H^{k,n}$ with $n\geq 2$ vanish.
The $\SO(6)\times \SO(4)$-invariant part is in $H^{0,0}$, and $H^{4,1}$.

Generators:
\beqr
 \left[t^{\alpha}t^{\beta}c_m \Gamma^{m}_{\alpha\beta}\right]\in H^{4,1}
\eeqr

Poincare series
\beqr
P_0(\tau)&=&\frac{4{\tau}^5 + 4{\tau}^4 - 5{\tau}^3 - 9{\tau}^2 - 5{\tau} - 1}{(-1 + {\tau})^{11}},\\
P_1(\tau)&=&\frac{{\tau}^4 + 5 {\tau}^5 + 5 {\tau}^6 + {\tau}^7}{(1 - {\tau})^{11}}
\eeqr

\item $r=5$,
\beq
H^{k,0}= \overset{\left[k/2\right]}{\underset{i=1}{\oplus}}\overset{i-1}{\underset{j=0}{\oplus}}[j,k-2i,i,k-2i] \overset{\left[k/2\right]}{\underset{i=0}{\oplus}}\overset{\left[(k-2i)/2\right]}{\underset{j=0}{\oplus}}[i,k-2i-2j,i,k-2i-2j]
\eeq
Groups $H^{k,n}$ with $n\geq 1$ vanish.
The $\SO(5)\times \SO(5)$-invariant part lies in  $H^{k,0}$ where $k$ is even.

Poincare series
\beq
P_0(\tau)=\frac{(1 + {\tau})^{5}}{(1 - {\tau})^{11}}
\eeq

\item $r=4$,
\beq
H^{k,0}= \overset{\left[k/2\right]}{\underset{i=0}{\oplus}}\overset{k-2i}{\underset{j=0}{\oplus}}(i+1)\times[j,k-2i-j,j,i,k-2i-j]
\eeq
where the coefficient $(i+1)$ is the multiplicity. Groups $H^{k,n}$ with $n\geq 1$ vanish.
The $\SO(4)\times \SO(6)$-invariant part is in $H^{0,0}$.

Poincare series
\beq
P_0(\tau)=\frac{(1 + {\tau})^{4}}{(1 -{\tau})^{12}}
\eeq

\item $r=3$,
\beq
H^{k,0}= \overset{\left[k/2\right]}{\underset{i=0}{\oplus}}\overset{i}{\underset{j=0}{\oplus}}[k-2i,j,i-j,k-2i]
\eeq
Groups $H^{k,n}$ with $n\geq 1$ vanish.
The $\SO(3)\times \SO(7)$-invariant part is in $H^{0,0}$.

Poincare series
\beq
P_0(\tau)=\frac{(1 + {\tau})^{3}}{(1 - {\tau})^{13}}
\eeq

\item $r=2$,
\beq
H^{k,0}= \overset{\left[k/2\right]}{\underset{i=0}{\oplus}}\overset{k-2i}{\underset{j=0}{\oplus}}[i,0,k-2i-j,j,k-2i-2j]
\eeq
Groups $H^{k,n}$ with $n\geq 1$ vanish.
The $\SO(2)\times \SO(8)$-invariant part is in $H^{0,0}$.

Poincare series
\beq
P_0(\tau)=\frac{(1 + {\tau})^{2}}{(1 - {\tau})^{14}}
\eeq

\item $r=1$,
\beq
H^{k,0}= \overset{\left[k/2\right]}{\underset{i=0}{\oplus}}[i,0,0,k-2i]
\eeq
Groups $H^{k,n}$ with $n\geq 1$ vanish.
The $\SO(1)\times \SO(9)$-invariant part is in $H^{0,0}$.

Poincare series
\beq
P_0(\tau)=\frac{1+{\tau}}{(1 - {\tau})^{15}}
\eeq

\end{itemize}

For $r\leq 5$, the cohomology is generated by the scalar $1$.

\section {Dimensional reduction from $D=11$\label{Sec:redux11D}}
Let us consider dimensional reductions of eleven-dimensional Lie algebra of supersymmetries. The reduction of $\susy _{11}$ to $r$ dimensions has $32$ odd generators (supersymmetries) and $r$ even generators (here $0\leq r\leq 11$). Corresponding differential has the form (\ref {sud} ) where  $\Gamma _{\alpha \beta}^m$ are eleven-dimensional Dirac matrices, Greek indices take $32$ values as in unreduced case, but Roman indices take only $d$ values. The differential commutes with action of the group $\SO(r)\times \SO (11-r)$, therefore this group acts on cohomology. The cohomology can be regarded as a module over $\mathbb{C}[t^1,...,t^{32}].$
Again cohomology is double graded; we use notation $H^{k,n}$ for the component having degree $m=k-2n$ with respect to $t$ and the degree $n$ with respect to $c$. The symbol $P_n(\tau)$ stands for the generating  function $P_n(\tau)=\sum_k \dim H^{k,n}\tau ^k$ (for Poincare series).
We calculate the cohomology as a representation of Lie algebra $\sso(r)\times \sso(11-r)$ and describe elements that generate it as a $\mathbb{C}[t^1,...,t^{32}]-$module.

Let us start  with calculation of Euler characteristic $\chi (H^{k})$ of cohomology $H^k=\sum _nH^{k,n}$. By general theorem this is a virtual $\sso(r)\times \sso(11-r)$-module
\begin{equation}
\label{eul}
 \sum _n(-1)^n\Sym ^{k-2n}S\otimes \Lambda ^nV
\end{equation}
where $S$ and $V$ are considered as $\sso(r)\times \sso(11-r)$-modules.

Cohomology for  $r=10$ are given by the formula
\beqr
H^{k,0}&=& \overset{\left[k/2\right]}{\underset{i=0}{\oplus}}\overset{k-2i}{\underset{j=0}{\oplus}}\overset{\left[\frac{k-2i-j}{2}\right]}{\underset{l=0,l\neq 2}{\oplus}}[0,i,0,j,k-2i-j-2l]\nonumber \\
&&\overset{\left[k/2\right]}{\underset{i=1}{\oplus}}\overset{\left[\frac{k-2i}{2}\right]}{\underset{j=0}{\oplus}}\overset{k-2i-2j}{\underset{l=0}{\oplus}}[i,j,0,l,k-2i-2j-l]\nonumber \\
&&\overset{\left[\frac{k-4}{2}\right]}{\underset{i=0}{\oplus}}\overset{k-4-2i}{\underset{j=0}{\oplus}}[0,i,0,j,k-4-2i-j], \\
H^{k,1}&=& \overset{\left[\frac{k-4}{2}\right]}{\underset{i=0}{\oplus}}\overset{i}{\underset{j=0}{\oplus}}\overset{k-4-2i}{\underset{l=0}{\oplus}}[j,i-j,0,l,k-4-2i-l] \\
\eeqr
Groups $H^{k,n}$ with $n\geq 2$ vanish.
The $\SO(10)$-invariant part is in $H^{0,0}$ and $H^{4,1}$.

Generators
\beq
 \left[t^{\alpha}t^{\beta}c_m \Gamma^{m}_{\alpha\beta}\right]\in H^{4,1}
\eeq

Poincare series
\beqr
P_0(\tau)&=&\frac{(1+{\tau})^{10}}{(1 - {\tau})^{22}}+{\tau}^4 A(\tau),\\
P_1(\tau)&=&{\tau}^4 A(\tau)
\eeqr
where $A(\tau)$ is the Poincare series given by Eq.~\ref{P11D:0}.

For $r\leq9$ the groups $H^{k,n}$ with $n\geq 1$ vanish hence Euler characteristic (\ref {eul}) gives a complete description of cohomology.

Poincare series
\beq
P_0(\tau)=\frac{(1 + {\tau})^{r}}{(1 - {\tau})^{32-r}}
\eeq

To find the $\sso(r)\times \sso(11-r)$- invariant part of $H^{k,0}$ it is sufficient to solve this problem for Euler characteristic. The conjectural answers (obtained by means of computations for $k<19$)
are listed below.

$r=1$,
\beq
 \left[0,0,0,0,0\right]\in H^{4k,0}
\eeq
$r=2$,
\beq
 (\left[k/2\right]+1)\times\left[0,0,0,0,0\right]\in H^{2k,0}
\eeq
$r=3$,
\beq
(k+1)\times \left[0,0,0,0,0\right]\in H^{4k,0}
\eeq
$r=4$,
\beq
\frac{(k+1)(k+2)}{2}\times \left[0,0,0,0,0\right]\in H^{4k,0}
\eeq
$r=5$,
\beq
\frac{(\left[k/2\right]+1)(\left[k/2\right]+2)(\left[k/2\right]+3)}{6}\times \left[0,0,0,0,0\right]\in H^{2k,0}
\eeq
$r=6$,
\beq
\frac{(\left[k/2\right]+1)(\left[k/2\right]+2)}{2}\times \left[0,0,0,0,0\right]\in H^{2k,0}
\eeq
$r=7$,
\beq
(k+1)\times \left[0,0,0,0,0\right]\in H^{2k,0}
\eeq
$r=8$,
\beq
(2k+1)\times \left[0,0,0,0,0\right]\in H^{2k,0}
\eeq
$r=9$,
\beqr
 2\times\left[0,0,0,0,0\right]\in H^{4k,0},\\
 \left[0,0,0,0,0\right]\in H^{4k+2,0}
\eeqr
 where $i\times[a,b,c,d,e]$ denotes the representation $[a,b,c,d,e]$ with multiplicity $i$ and $[a]$ stands for the integer part of $a$.

\section {Other dimensions\label{Sec:otherD}}

In this section we consider in detail cohomology of Lie algebra of supersymmetries in  dimensions $<10$. Let us begin with some general discussion
 of supersymmetries in various dimensions (see \cite {Deligne} and \cite {M} for more detail).

We will work   with complex Lie algebras. Let us start with the description of the symmetric intertwiners
$\Gamma: S^*\otimes S^*\to V$ used in the construction of supersymmetry Lie algebra in various dimensions (notice that in the construction of differential we use dual intertwiners). Recall that in even dimensions we have two irreducible spinorial representations $\s_l$ and $\s_r$, in odd dimensions we have one irreducible spinorial representation $\s$.
\begin{itemize}
\item $\dim V=8n$

In this case we have intertwiners $\gamma _l:\s_l\otimes \s_r\to V$ and $ \gamma _r:\s_r\otimes \s_l\to V$.
$$ S\cong S^*=\s_l+\s_r,\quad \Gamma=\gamma _l+\gamma _r,\quad \dim S=16^n.$$
Automorphism Lie algebra $aut=\sso (8n)\times \sso (2).$

\item $\dim V=8n+1$

In this case we have symmetric intertwiner $\gamma:\s\otimes\s\to V.$
$$S=S^*=\s,\quad \Gamma=\gamma,\quad \dim S=16^n.$$
Automorphism Lie algebra $aut=\sso (8n+1).$

\item $\dim V=8n+2$

In this case we have symmetric intertwiners $\gamma _l:\s_l\otimes \s_l\to V$ and  $\gamma _r:\s_r\otimes \s_r\to V.$

There are two possible choices of $S:$
$$S=\s_r, S^*=\s_l,\Gamma=\gamma_l;\quad S=\s_l, S^*=\s_r,\Gamma=\gamma_r, \dim S=16^n.$$
Automorphism Lie algebra $aut=\sso (8n+2 ).$

\item $\dim V=8n+3$

In this case we have symmetric intertwiner $\gamma:\s\otimes\s\to V.$
$$S=S^*=\s,\quad \Gamma=\gamma,\quad \dim S=2\times 16^n.$$
Automorphism Lie algebra $aut=\sso (8n+3).$

\item $\dim V=8n+4$

In this case we have intertwiners $\gamma _l:\s_l\otimes \s_r\to V$ and $ \gamma _r:\s_r\otimes \s_l\to V.$
$$ S\cong S^*=\s_l+\s_r,\quad \Gamma=\gamma _l+\gamma _r,\quad \dim S=4\times 16^n.$$
Automorphism Lie algebra $aut=\sso (8n+4)\times \sso (2).$

\item $\dim V=8n+5$

The intertwiner $\gamma: \s\otimes \s \to V$ is antisymmetric.
$$S\cong S^*=\s\otimes W,\Gamma=\gamma\otimes \omega, \dim S=8\times 16^n.$$
Here and later $W$ stand for two-dimensional linear space with a symplectic form $\omega$.
Automorphism Lie algebra $aut=\sso (8n+5)\times \sl(2).$

\item $\dim V=8n+6$

In this case we have antisymmetric intertwiners $\gamma _l:\s_l\otimes \s_l\to V$ and  $\gamma _r:\s_r\otimes \s_r\to V.$
There are two possible choices of $S:$
$$S^*=\s_l\otimes W,\Gamma=\gamma_l\otimes \omega;\quad S^*=\s_r\otimes W,\Gamma=\gamma_r\otimes \omega, \dim S=8\times 16^n.$$
Automorphism Lie algebra $aut=\sso (8n+6)\times \sl(2).$

\item $\dim V=8n+7$

The intertwiner $\gamma: \s\otimes \s \to V$ is antisymmetric.
$$S=S^*=\s\otimes W,\Gamma=\gamma\otimes \omega, \dim S=16\times 16^n.$$

Automorphism Lie algebra $aut=\sso (8n+7)\times \sl(2).$
\end {itemize}

One can consider also $N$-extended supersymmetry Lie algebra. This means that we  should start with reducible spinorial representation $S_N$(direct sum of $N$ copies of  spinorial representation $S$). Taking $N$ copies of the intertwiner  $V\to \Sym^2S$ we obtain an intertwiner $V\to \Sym^2S_N$. We define the $N$-extended supersymmetry Lie algebra by means of this intertwiner. The Lie algebra acting on its cohomology acquires an additional factor  $\gl(N)$ .

Notice that in the cases when there are two different possible choices of $S$ (denoted by $S_1$ and $S_2$) one can talk about
$(N_1,N_2)$-extended supersymmetry taking  as a starting point a direct sum  of $N_1$ copies of $S_1$ and $N_2$ copies of $S_2.$

The description of cohomology of supersymmetry Lie algebras in dimensions 9,8,7  follows immediately from the description of cohomology of  ten-dimensional supersymmetry Lie algebra  reduced to these dimensions. (Notice $S$ has dimension $16$ in all of these cases.)

We will describe the cohomology of the Lie algebra of  supersymmetries in six-dimensional case as representations of the Lie algebra $\sso(6)\times \sl(2)$. The vector representation $V$ of $\sso(6)$ has the highest weight $[1,0,0]$, the irreducible spinor representations have highest weights $[0,0,1]$, $[0,1,0]$; we consider for definiteness $\s_l$ with highest weight $[0,0,1]$. As a representation $\sso(6)\times \sl(2)$ the representation $V$ has the weight $[1,0,0,0]$ and the representation $S=\s_l\otimes W$ has the weight $[0,0,1,1]$. The description
of graded component of cohomology group with gradings $k=m+2n$ and $n$ is given by the formulas
\beqr
H^{k,0}&=& [0,0,k,k] \label{Hk6D,0}\\
H^{k,1}&=& [1,0,k-3,k-2] \\
H^{k,2}&=& [0,1,k-6,k-4] \\
H^{k,3}&=& [0,0,k-8,k-6] \label{Hk6D,3}
\eeqr
The only special case is when $k=4$, there is one additional term, a scalar, for $H^{4,1}$.
\beq
H^{4,1}=[0,0,0,0]\oplus[1,0,1,2] \label{H46D,1}
\eeq
 For $n\geq 4$, $H^{k,n}$ vanishes.
 The $\sso(6)\times \sl(2)$-invariant part is in $H^{0,0}$, and $H^{4,1}$.

The dimensions of the cohomology groups are encoded in Poincare series:
\beqr
P_0(\tau)&=&\frac{1 + 3{\tau}}{(1 - {\tau})^{5}}, \\
P_1(\tau)&=&\frac{8{\tau}^3 + 6{\tau}^4 - 6{\tau}^5 + 10{\tau}^6 - 10{\tau}^7 + 5{\tau}^8 -{\tau}^9}{(1 - {\tau})^{5}}, \\
P_2(\tau)&=&\frac{18{\tau}^6 - 10{\tau}^7 + 5{\tau}^8 -{\tau}^9}{(1 - {\tau})^{5}},\\
P_3(\tau)&=&\frac{3{\tau}^8 +{\tau}^9}{(1 - {\tau})^{5}}
\eeqr

The cohomology considered as ${\C}[t^1,..., t^{\alpha},...]$-module is generated by the scalar and
\beqr
 \left[t^{\alpha}c_m \Gamma^m_{\alpha\beta}\right]\in H^{3,1}, \nonumber\\
 \left[t^{\alpha}t^{\beta}c_m c_n \Gamma^{mnk}_{\alpha\beta}\right]\in H^{6,2}, \nonumber\\
 \left[t^{\alpha}t^{\beta}c_m c_n c_k \Gamma^{mnk}_{\alpha\beta}\right]\in H^{8,3} \nonumber
\eeqr

Now we will describe the cohomology of the Lie algebra of  supersymmetries in five-dimensional case as representations of the Lie algebra $\sso(5)\times \sl(2)$. The vector representation $V$ of $\sso(5)$ has the highest weight $[1,0]$, the irreducible spinorial representation has highest weight $[0,1]$. As a representation $\sso(5)\times \sl(2)$ the representation $V$ has the weight $[1,0,0]$ and the representation $S=\s\otimes W$ has the weight $[0,1,1]$. The description
of graded component of cohomology group with gradings $k=m+2n$ and $n$ is given by the following formulas  (for $n\geq 3$, $H^{k,n}$ vanishes)
\beqr
H^{k,0}&=& [0,k,k] \label{Hk5D,0}\\
H^{k,1}&=& [1,k-4,k-2] \\
H^{k,2}&=& [0,k-6,k-4] \label{Hk5D,3}
\eeqr
The only special case is when $k=2$, there is one additional term, a scalar, for $H^{2,1}$.
\beq
H^{2,1}=[0,0,0]\oplus[0,2,2] \label{H2D5,1}
\eeq
The $\sso(5)\times \sl(2)$-invariant part is in $H^{0,0}$, and $H^{2,1}$.

The dimensions of the cohomology groups are encoded in Poincare series:
\beqr
P_0(\tau)&=&\frac{1 + 3{\tau} + {\tau}^2  - 5{\tau}^3  + 10{\tau}^4  - 10{\tau}^5  + 5{\tau}^6  - {\tau}^7}{(1 - {\tau})^{5}}, \\
P_1(\tau)&=&\frac{15{\tau}^4  - 11{\tau}^5  + 5{\tau}^6  - {\tau}^7}{(1 - {\tau})^{5}}, \\
P_2(\tau)&=&\frac{3{\tau}^6  + {\tau}^7}{(1 - {\tau})^{5}}
\eeqr

The cohomology regarded as ${\C}[t^1,..., t^{\alpha},...]$-module is generated by the scalar considered as an element of $H^{0,0}$ and
\beqr
 \left[t^{\alpha}t^{\beta}c_m \Gamma^{mn}_{\alpha\beta}\right]\in H^{4,1}, \nonumber\\
 \left[t^{\alpha}t^{\beta}c_m c_n \Gamma^{mn}_{\alpha\beta}\right]\in H^{6,2} \nonumber
\eeqr

In four-dimensional case  the  representation $S$
should be considered as  4-dimensional Dirac spinor.

We  describe the cohomology of the Lie algebra of  supersymmetries in four-dimensional case as representations of the Lie algebra $\sso(4)$. As usual the representations are labeled by their
highest weight. The vector representation $V$ has the highest weight $[1,1]$, the irreducible spinor representations have highest weights $\s_l=[0,1]$, $\s_r=[1,0]$; we assume that $S=\s_l+\s_r=[0,1]\oplus[1,0]$. The description
of graded component of cohomology group with gradings $k=m+2n$ and $n$ is given by the  following formulas  (for $n\geq 6$, $H^{k,n}$ vanishes)
\beqr
H^{k,0}&=& [0,k]\oplus [k,0] \label{Hk4,0}\\
H^{k,1}&=& [1,k-3]\oplus [k-3,1] \\
H^{k,2}&=& [0,k-6]\oplus [k-6,0] \label{Hk4},
\eeqr

The only special case is when $k=4$, there is one additional term, a scalar, for $H^{4,1}$.
\beq
H^{4,1}=[0,0]\oplus 2\times[1,1] \label{H4D4,1}
\eeq
The $\sso(4)$-invariant part is in $H^{0,0}$, $H^{6,2}$, and $H^{4,1}$.

The dimensions of these cohomology groups are encoded in Poincare series:
\beqr
P_0(\tau)&=&\frac{1+2{\tau}- {\tau}^2}{(1 - {\tau})^{2}}, \\
P_1(\tau)&=&\frac{4{\tau}^3  + {\tau}^4  - 2{\tau}^5  + {\tau}^6}{(1 - {\tau})^{2}}, \\
P_2(\tau)&=&\frac{2{\tau}^6}{(1 - {\tau})^{2}}
\eeqr

The cohomology can be regarded as ${\C}[ t^{\alpha},t^{\dot{\alpha}}]$-module
where $\alpha=1,2$, $\dot{\alpha}=\dot{1},\dot{2}$. (Here $t^{\alpha}$ transforms according to the representation $[1,0]$ and $t^{\dot{\alpha}}$ transforms according to the representation $[0,1].$) This module
 is generated by the scalar considered as an element of $H^{0,0}$ and
\beqr
 \left[t^{\alpha}c_m \Gamma^{m}_{\alpha\dot{\beta}}\right]\in H^{3,1}, \nonumber\\
 \left[t^{\dot{\alpha}}c_m \Gamma^{m}_{\dot{\alpha}{\beta}}\right]\in H^{3,1}, \nonumber\\
 \left[t^{\alpha}t^{\dot{\beta}}c_m c_n \Gamma^{mn}_{\alpha\dot{\beta}}\right]\in H^{6,2}, \nonumber\\
 \left[t^{\dot{\alpha}}t^{{\beta}}c_m c_n \Gamma^{mn}_{\dot{\alpha}{\beta}}\right]\in H^{6,2} \nonumber
\eeqr

The cohomology generators in  $D=4$ and $D=5$ were found by F. Brandt~\cite{Brandt}.

\section{Calculations\label{Sec:Calculations}}
We  do calculation of Poincare series applying \cite {mac2}. However, the straightforward calculation is pretty lengthy with the computers we are using. Therefore for $D=10$ and $D=11$
we consider dimensional reduction to dimension $r$ and we are using induction with respect to $r$.

Recall that the differential  of $D$-dimensional theory reduced to dimension $r$ has the form

\begin {equation}
\label {sur}
d_r=\sum _{1\leq m\leq r}A^m\frac{\partial}{\partial c^m}.
\end {equation}
where $A^m=\frac{1}{2}\Gamma _{\alpha \beta}^mt^{\alpha}t^{\beta}$
and acts in the space $E_r$ of polynomial functions of even ghosts $t ^{\alpha}$ and $r$ odd ghosts $c^1,...,c^r$.  We denote corresponding cohomology by $H_r.$ Both $E_r$ and $H_r$ are bigraded  by the degree of even ghosts $m$ and degree of odd ghosts $n$, but it is simpler to work with  grading with respect to $k=m+2n$ and $n$. An element of $E_r$ can be represented in the form $x+yc^r$ where $x,y\in E_{r-1}.$ Notice that
$$d_r(x+yc^r)=d_{r-1}x+A^ry+d_{r-1}yc^r.$$

The operator of  multiplication by $A_m$ commutes with the differential, hence it induces a homomorphism
$$\sigma :H_{r-1}\to H_{r-1}.$$
Sending $x\in E_{r-1}$ into $x+0c^m\in E_r$  (embedding $E_{r-1}$ into $E_r$)  we obtain a homomorphism $H_{r-1}\to H_r$. Sending $x+yc^m$ into $y$ we get a homomorphism $H_r\to H_{r-1}.$ It is easy to see that  combining these homomorphisms we obtain an exact sequence
$$H_{r-1}\to H_{r-1} \to H_r\to H_{r-1} \to H_{r-1}$$
or, taking into account the gradings,
\begin{equation}
\label{ex}
\cdots \to H_{r-1}^{k,n}\to H_{r-1}^{k+2,n} \to H_r^{k+2,n}\to H_{r-1} ^{k,n-1}\to H_{r-1}^{k+2,n-1}\to \cdots
\end{equation}
(This is the exact sequence of a pair $(E_r,E_{r-1})$; we use the fact that $E_r^{k,n}/E_{r-1}^{k,n}\cong E_{r-1}^{k-2,n-1}$.) It follows immediately from this exact sequence that an  isomorphism $H_{r-1}^{k,n}=0$ for $n>n_{r-1}$ implies $H_r^{k,n}=0$ for $n>n_r+1$. (In other  words if $n_r$ is the maximal degree of cohomology in $r$-dimensional reduction then $n_{r+1}\leq n_r+1$.) Applying the exact sequence (\ref {ex}) to the case $n=n_r$ and assuming that $n_{r-1}<n_r$ we obtain an isomorphism
between $H_r^{k+2,n_r}$ and a subgroup of $H_{r-1} ^{k,n_{r-1}}$ (this isomorphism can be considered as an isomorphism of $\sso (r-1)$-representations).  For dimensional reductions of $D=10$ and $D=11$ algebras of supersymmetries  dimensions of  $H_r^{k+2,n_r}$ and  $H_{r-1} ^{k,n_{r-1}}$ coincide because Poincare series are related by the formula $P_{n_r}=\tau ^2 P_{n_{r-1}}$.
If homomorphism  $\sigma$ is injective we obtain a short exact sequence
$$0 \to H_{r-1}^{k,n}\to H_{r-1}^{k+2,n} \to H_r^{k+2,n}\to 0.$$
 Calculations with \cite {mac2} show that $n_1=...=n_5=0$ for $D=10$ and $n_1=...=n_9=0$ for $D=11$. (It is sufficient to check that in corresponding dimensions the homomorphism $\sigma$ is injective.)

 To analyze $r$-dimensional reduction  for $r>5, D=10$  we notice that $d_r$ can be considered as a sum of  differentials $d'$ and $d''$ where
 $$d'=\sum _{1\leq m\leq 5}A^m\frac{\partial}{\partial c^m},$$
 $$d''=\sum _{5< m\leq r}A^m\frac{\partial}{\partial c^m}.$$
 (For  $r>9, D=11$  one should replace $5$ by $9$.)
 These differentials anticommute; this allows us to use the spectral  sequence of bicomplex to calculate the cohomology of $d_r$. The spectral sequence of bicomplex starts with cohomology
 $H(d'', H(d'))$. Taking into account  that the cohomology $H(d')=H(d_5)$ is concentrated in degree $0$ (as the cohomology $H_5$)  we obtain that the spectral sequence terminates. This means that one can calculate the Poincare series of $d_r$ as   the Poincare series of $H(d'', H(d'))$ using \cite {mac2}. Again applying \cite {mac2} we can obtain the information about generators of cohomology; this information is sufficient to express the generators in terms of Gamma-matrices.

To calculate the cohomology as a representation of the group of automorphisms we decompose each graded component $E^{k,n}=\Sym^{k-2n}S\otimes \Lambda ^n V$ of $E$ into direct sum of irreducible representations.

For example, for $D=10$ spacetime, we have the cochain complex
\beq
\begin{split}
0& \xleftarrow{d_0} \Sym^kS \xleftarrow{d_1} \Sym^{k-2}S\otimes V \xleftarrow{d_2} \Sym^{k-4}S\otimes \wedge^2 V \\
&\xleftarrow{d_3} \Sym^{k-6}S\otimes \wedge^3 V \xleftarrow{d_4} \Sym^{k-8}S\otimes \wedge^4V \xleftarrow{d_5}\Sym^{k-10}S\otimes \wedge^5V \\
&\xleftarrow{d_6}\Sym^{k-12}S\otimes \wedge^6V  \xleftarrow{d_7}\Sym^{k-14}S\otimes \wedge^7V \xleftarrow{d_8}\Sym^{k-16}S\otimes \wedge^8V \\
&\xleftarrow{d_9}\Sym^{k-18}S\otimes \wedge^9V \xleftarrow{d_{10}}\Sym^{k-20}S\otimes \wedge^{10}V \xleftarrow{d_{11}} 0 \label{complex.10D}
\end{split}
\eeq
where for $\Sym^mS\otimes \wedge^nV$, the  grading index $k=m+2n$ is preserved by $d$. All components of this complex can be regarded as representations of $\sso(10)$. We have
\beq
\begin{split}
&S=[0,0,0,0,1] \text{ (choosen) or } [0,0,0,1,0], \quad V=[1, 0,0,0,0] \\
&\wedge^2V=[0,1,0,0,0], \quad \wedge^3V=[0,0,1,0,0],    \\
&\wedge^4V=[0,0,0,1,1], \quad \wedge^5V=[0,0,0,0,2]\oplus [0,0,0,2,0],    \\
&\wedge^6V=\wedge^4V, \quad \wedge^7V=\wedge^3V, \quad \wedge^8V=\wedge^2V, \quad \wedge^9V=V, \quad \wedge^{10}V=[0,0,0,0,0],   \\
\end{split}
\eeq

\beq
\Sym^kS=
\overset{\left[k/2\right]}{\underset{i=0}{\oplus}}[i,0,0,0,k-2i]  \label{SymkS_10D}
\eeq
(see \cite {MSX} for the decomposition of $\Sym^m S\otimes \wedge^n V$ and for complete description of action of differential on irreducible components for supersymmetry Lie algebra in 10D
and 6D.)

By the Schur's lemma an intertwiner between irreducible representations (a homomorphism of simple modules) is either zero or an isomorphism. This means that an intertwiner between non-equivalent
irreducible representations  always vanishes. This observation permits us to calculate the contribution of every irreducible representation to the cohomology separately.

Let us fix an irreducible representation $A$ and the number $k$. We will denote by $\nu _n$ (or by $\nu_n^k$ if it is necessary to show the dependence of $k$) the multiplicity of $A$ in $E^{k,n}=\Sym^{k-2n}S\otimes \Lambda ^n V.$ The multiplicity of $A$ in the image of
$d:E^{k,n}\to E^{k,{n-1}}$ will be denoted by $\kappa _n$,
  then the multiplicity of $A$ in the kernel of this map is equal to $\nu _n-\kappa _n$ and the multiplicity of $A$ in the cohomology $H^{k,n}$
is equal to $h_n=\nu _n-\kappa _n-\kappa _{n+1}.$ It follows immediately that the multiplicity of $A$ in virtual representation $\sum _n (-1)^n H^{k,n}$ (in the Euler characteristic) is equal to $\sum
(-1)^n\nu _n.$  It does not depend on $\kappa _n$, however, to calculate the cohomology completely we should know $\kappa _n$.

Let us consider as an example $A=[0,0,0,0,0]$, the scalar representation, for dimension $D=10$ and arbitrary $k$.  For all $k\neq 4,12$, we have $\nu_i=0$ .  (For small $k$ this can be obtained by means of LiE program \cite {LiEcode}.) For $k=4$, we have all $\nu_i$ vanish except $\nu_1=1$, hence all $\kappa_i$ vanish. The multiplicity of $[0,0,0,0,0]$ in $H^{4,1}$ is equal to $1$, and other cohomology $H^{4,i}$ do not contain scalar representation. For $k=12$, all $\nu_i$ vanish except $\nu_5=1$, hence $H^{12,5}$ contains $[0,0,0,0,0]$ with multiplicity $1$, and $H^{12,i}$ do not contain $[0,0,0,0,0]$ for $i\neq 5$. This agrees with Eq.~\ref{H4,1} and Eq.~\ref{Hk,5}, respectively.

In many cases a heuristic calculation of cohomology can be based on a
principle that kernel should be as small as possible; in other words, the image should be as large as possible (this is an analog of the general rule of the physics of elementary particles: Everything
happens unless it is forbidden). In \cite{Bengt.Nilsson} this is called the principle of maximal propagation. {\footnote { Notice that the principle of maximal propagation should be applied to the decomposition of cohomology into irreducible representations of the full automorphism group. Otherwise we do not use all available information.}}
Let us illustrate this principle in the case when $k=9$ and $A=[0,1,0,0,1]$ in $10D$.
In this case $\nu_4=1$, $\nu_3=3$, $\nu_2=1$. If we believe in the maximal propagation, then $\kappa_3=1,\kappa_4=1$, thus we have $\nu_3 - \kappa_3 - \kappa_4 =1$, and $[0,1,0,0,1] $ contributes only to $H^{9,3}$.

Notice, that the principle of maximal propagation sometimes does not give a definite answer. For example, in the case when $k=8$ in the dimension reduced to $7$ from $10D$. Considering only the multiplicities of $A=[0,0,2,2]$,
we have sequence $0\to 0\to 2\to 5\to 3\to 1$. This sequence
offers two distinct possibilities even under the assumption of maximal propagation. We can assume that the kernels of the differentials $2\to 5$ and $5\to 3$ are minimal. In this case $h_0=1$  .    Or we can start with assumption that the kernel  of differential $3\to 1$ is minimal, then the kernel of  $5\to 3$ has multiplicity at least $3$ and assuming that this multiplicity is equal to $3$ we see under the assumption of minimality of the kernel of $2\to 5$ that the only non-trivial cohomology is  $h_2=1$. (We  can  prove that the second position is the correct choice.)  We see that the result of the method of maximal propagation depends on what differential we choose to start with.  More generally, we should  order  the differentials in the complex in some way and apply the principle of maximal propagation using this ordering.

In  cases we are interested in one can prove that the principle of maximal propagation augmented with  information about Poincare series and generators  is free of ambiguities.  In other words, this information permits us to resolve ambiguities in the application of this principle.
Sometimes it is useful to apply the remark that  multiplying a coboundary by a polynomial we again obtain a coboundary.

The only exception is the case of ten-dimensional reduction of eleven-dimensional supersymmetry Lie algebra. In this case we use the isomorphism  between $H_{10}^{k,1}$ and $H_{11}^{k-2,2}$ that was derived from Eq.~\ref{ex}. This is an isomorphism of $\sso (10)$-representations; it allows us to find the decomposition of  $H_{10}^{k,1}$ from decomposition of  $H_{11}^{k-2,2}$ in irreducible representations of $\sso (11)$.  From the other side we can find the virtual  $\sso (10)$-character of $H_{10}^{k,0}-H_{10}^{k,1}$ (Euler characteristic); this allows us to finish the calculation.

Let $\cdots \to M_n\to \cdots \to M_0\to M\to 0$ denote the minimal free  $\sum \Sym ^m S$-resolution of the module $M=\sum _k H^{k,n}$. Then every free module $M_i$ in this resolution is  a representation of the group of automorphisms $Aut$ ; it is  a tensor product of a finite dimensional graded representation  $\mu _i$ (the  generators ) and a representation of $Aut$ in  polynomial algebra $\sum \Sym ^m S$. It is easy to find the dimensions of $Aut$-modules $\mu _i$ (the number of generators of $M_i$) using \cite {mac2}.
Dimensions of the graded components of  $\mu _i$ can be found routinely using \cite {mac2}.

The information about  free resolution can be used to find the structure of
$Aut$ -module on $\mu _i$ and therefore on $M$. However, we  went in opposite direction: we used the information about the structure  of $Aut$ -module on $M$ to find the structure of $Aut$ -module on $\mu _i$ using the formula {\footnote {This formula follows from well known $K$-theory relation $$(\sum \tau ^m \Sym ^mS )(\sum (-\tau)^j  \Lambda ^jS)=1.$$Taking into account  that parity reversal transforms symmetric power into exterior power we can understand this relation in the framework of super algebra.}}
\beq
\label{Eq:alt_mu}
 \sum_i(-1)^i\mu _i=(\sum _k H^{k,n}\tau ^k)\otimes(\sum (-\tau)^j  \Lambda ^jS).
\eeq
 The analysis of the resolution of the cohomology module is relegated to the appendix.

\section{Homology of super Poincare Lie algebra\label{Sec:superPoincare}}

The super Poincare Lie algebra can be defined as super Lie algebra spanned by supersymmetry Lie algebra and Lie algebra $aut$ of its group of automorphisms $Aut$. {\footnote {Instead of Lie algebra of automorphisms one can take its subalgebra. For example, we can take as a subalgebra the orthogonal Lie algebra }}

  To calculate the homology and cohomology of super Poincare Lie algebra we will use the following statement proved by Hochschild and Serre \cite{HochSerre} .(It follows from
  Hochschild-Serre spectral sequence constructed in the same paper.)

Let $\cal P$ denote a Lie algebra represented as a vector space as a direct sum of two subspaces $\cal L$ and $\cal G$. We assume that $\cal G$ is an ideal in $\cal P$ and that $\cal L$ is semisimple.
It follows from the assumption that $\cal G$ is an ideal that $\cal L$ acts on $\cal G$ and therefore on cohomology of $\cal G$; the $\cal L$-invariant part of cohomology $H^{\bullet}(\cal G))$ will
be denoted by $H^{\bullet}(\cal G))^{\cal L}$. One can prove that
$$H^n({\cal P})=\sum_{p+q=n}H^p({\cal L})\otimes H^q(\cal G)^{\cal L}.$$

This statement remains correct if $\cal P$ is a super Lie algebra. We will apply it to the case when $\cal P$ is super Poincare Lie algebra, $\cal  G$ is the Lie algebra of supersymmetries and $\cal L$ is the Lie algebra of automorphisms or its semisimple subalgebra.
(We are working with complex Lie algebras, but we can work with their real forms. The results do not change .)

Notice that it is easy to calculate the cohomology of semisimple Lie algebra $\cal L$; they are described by antisymmetric tensors on $\cal L$ that are invariant with respect to adjoint
representation. One can say also that they coincide with de Rham cohomology of corresponding compact  Lie group. The Lie algebra cohomology of $L=\sso_{10}$ with trivial coefficients  and as well as De Rham cohomology of  the compact Lie group  $\SO(10, \mathbb{R})$  is a Grassmann algebra with generators of dimension 3,7,11,13 and 9. In general the cohomology of the group $\SO(2r, \mathbb{R})$ is a Grassmann algebra with generators $e_i$ having dimension
$4i-1$ for $i<r$ and the dimension $2r-1$ for $i=r$. The cohomology of the group $\SO(2r +1, \mathbb{R})$ is a Grassmann algebra with generators $e_i$ having dimension
$4i-1$ for $i\leq r$ . The cohomology of Lie algebra $\sl(n)$ coincide with the cohomology of compact Lie group $\SU(n)$; they form a Grassmann algebra with
generators of dimension $3, 5,...,2n-1.$

As we have seen only $\cal L$-invariant part of cohomology of Lie algebra of supersymmetries
contributes to the cohomology of super Poincare algebra. For $D=10$ this means that the only contribution comes from subspaces $\Sym^m S\otimes \Lambda ^n V$ having the following indices $(m,n)=(0,0), (m,n)=(2,1)$ and $(m,n)=(2,5)$, for $D=11$ the only contribution comes from $(m,n)=(0,0)$ and $(m,n)=(2,2)$, for $D=6$ the only contribution comes from  $(m,n)=(0,0)$ and $(m,n)=(2,1).$ (Here $m=k-2n$ denotes the grading with respect to even ghosts $t^{\alpha}$ and $n$ the grading with respect to odd ghosts $c_m.$)

Cocycles representing cohomology classes of super Poincare
algebra can be written in the form $\rho\otimes h$, where $\rho$ is an  invariant antisymmetric tensor with respect to adjoint
representation of $aut$ and $h$ is
$1$ or
\beq
t^{\alpha}t^{\beta}c_m c_n \Gamma^{mn}_{\alpha\beta} \text{ for $D=11$}
\eeq
\beq
t^{\alpha}t^{\beta}c_m \Gamma^m_{\alpha\beta},\quad t^{\alpha}t^{\beta}c_m c_n c_k c_l c_r \Gamma^{mnklr}_{\alpha\beta} \text{ for $D=10$}
\eeq
\beq
t^{\alpha}t^{\alpha},\quad t^{\alpha}t^{\beta}c_m c_n c_k c_l \Gamma^{mnkl}_{\alpha\beta} \text{ for $D=9$}
\eeq
\beq
t^{\alpha}t^{\beta}c_m c_n c_k \Gamma^{mnk}_{\alpha\beta} \text{ for $D=8$}
\eeq
\beq
t^{\alpha}t^{\beta}c_m c_n \Gamma^{mn}_{\alpha\beta} \text{ for $D=7$}
\eeq
\beq
t^{\alpha}t^{\beta}c_m \Gamma^m_{\alpha\beta},\quad t^{\alpha}t^{\beta}c_m c_n c_k \Gamma^{mnk}_{\alpha\beta} \text{ for $D=6$}
\eeq
\beq
t^{\alpha}t^{\beta}c_m c_n \Gamma^{mn}_{\alpha\beta} \text{ for $D=5$}
\eeq
\beq
 t^{\alpha}t^{\dot{\beta}}c_m \Gamma^{m}_{\alpha\dot{\beta}},t^{\dot{\alpha}}t^{{\beta}}c_m \Gamma^{m}_{\dot{\alpha}{\beta}},t^{\alpha}t^{\dot{\beta}}c_m c_n \Gamma^{mn}_{\alpha\dot{\beta}},t^{\dot{\alpha}}t^{{\beta}}c_m c_n \Gamma^{mn}_{\dot{\alpha}{\beta}} \text{ for $D=4.$}
\eeq
Here Greek indices (i.e. spinor indices) take values $1,2,\cdots,\dim S$ and Roman indices (i.e. vector indices) take values $1,2,\cdots,D$, and $\dim S$ is defined in Section.\ref{Sec:otherD}. The only exception is for $D=4$, the Greek indices $\alpha,\beta$ take values $1,2$, and the dotted Greek indices $\dot{\alpha},\dot{\beta}$ take values $\dot{1},\dot{2}$. Notice that in these formulas Gamma matrices and summation range depend on the choice of dimension.

The general definition of  super Poincare algebra can be applied also to reduced supersymmetry Lie algebra. For $D=10$ and $D=11$  the role of super Poincare Lie algebra is played by the semidirect product of  reduced supersymmetry Lie algebra and $\sso(r)\times \sso (D-r).$  The information about invariant elements provided in Sections \ref{Sec:redux10D} and \ref{Sec:redux11D} permits us to describe cohomology of this generalization of super Poincare algebra.

\appendix

\section{Resolution of the cohomology modules\label{Appendix:resolution}}

One can find a minimal  free resolution of the $R$-module $\sum _k H^{k,n}=M$. (Here  $R=\mathbb{C}[t^1,\cdots,t^{\alpha},\cdots]=\sum_m \Sym^m S.$) The reader may wish to consult  \cite{CE} on this subject. The free resolution has the form
$$\cdots \to M_i\to \cdots \to M_0\to M\to 0$$
where $M_i=\mu_i\otimes R,$ and \\
$\mu_0$ - generators of $M$;\\
$\mu_1$ - relations between generators of $M$;\\
$\mu_2$ - relations between relations ;\\
$\cdots$

\subsection{Resolution of the cohomology modules of dimensional reduction of 10D Lie algebra of supersymmetries}
We give the structure of $\mu_i$ as $aut$-module and its grading in the case of $r$-dimensional reduction of ten-dimensional Lie algebra of supersymmetries  (in the case when $D=r+(10-r)$). Recall that $aut$ denotes the Lie algebra of the group of automorphisms $Aut$.

\begin{itemize}
\item D=10+0, n=0
$$\mu_0=[0,0,0,0,0], \dim(\mu_0)=1, \deg(\mu_0)=0; $$
$$\mu_1=[1,0,0,0,0], \dim(\mu_1)=10, \deg(\mu_1)=2; $$
$$\mu_2=[0,0,0,1,0], \dim(\mu_2)=16, \deg(\mu_2)=3; $$
$$\mu_3=[0,0,0,0,1], \dim(\mu_3)=16, \deg(\mu_3)=5; $$
$$\mu_4=[1,0,0,0,0], \dim(\mu_4)=10, \deg(\mu_4)=6; $$
$$\mu_5=[0,0,0,0,0], \dim(\mu_5)=1, \deg(\mu_5)=8. $$

\item D=10+0, n=1
$$\mu_0=[0,0,0,1,0], \dim(\mu_0)=16, \deg(\mu_0)=3; $$
$$\mu_1={\mu_1}'+{\mu_1}'', $$
$${\mu_1}'=[0,1,0,0,0], \dim({\mu_1}')=45, \deg({\mu_1}')=4; $$
$${\mu_1}''=[0,0,0,0,1], \dim({\mu_1}'')=16, \deg({\mu_1}'')=5; $$
$$\mu_2=2\times[0,0,1,0,0] +[1,0,0,0,0], \dim(\mu_2)=250, \deg(\mu_2)=6; $$
$$\mu_3=[0,0,0,1,0] +[0,1,0,1,0] +[1,0,0,0,1], \dim(\mu_3)=720, \deg(\mu_3)=7; $$
$$\mu_4={\mu_4}'+{\mu_4}'', $$
$${\mu_4}'=[0,2,0,0,0] +[1,0,0,2,0] +[2,0,0,0,0], \dim({\mu_4}')=1874, \deg({\mu_4}')=8; $$
$${\mu_4}''=[0,0,0,0,1], \dim({\mu_4}'')=16, \deg({\mu_4}'')=9; $$
$$\mu_5={\mu_5}'+{\mu_5}'', $$
$${\mu_5}'=[0,0,0,0,1] +[0,0,0,0,1] +[0,0,0,3,0] +[1,1,0,1,0], \dim({\mu_5}')=4352, \deg({\mu_5}')=9; $$
$${\mu_5}''=[1,0,0,0,0], \dim({\mu_5}'')=9, \deg({\mu_5}'')=10; $$
$$\mu_6=[0,1,0,2,0] + [2,0,1,0,0], \dim(\mu_6)=8008, \deg(\mu_6)=10; $$
$$\mu_7=[1,0,1,1,0] +[3,0,0,0,1], \dim(\mu_7)=11440, \deg(\mu_7)=11; $$
$$\mu_8=[0,0,2,0,0] +[2,0,0,1,1] +[4,0,0,0,0], \dim(\mu_8)=12870, \deg(\mu_8)=12; $$
$$\mu_9=[1,0,1,0,1] +[3,0,0,1,0], \dim(\mu_9)=11440, \deg(\mu_9)=13; $$
$$\mu_{10}=[0,1,0,0,2] +[2,0,1,0,0], \dim(\mu_{10})=8008, \deg(\mu_{10})=14; $$
$$\mu_{11}=[0,0,0,0,3] +[1,1,0,0,1], \dim(\mu_{11})=4368, \deg(\mu_{11})=15; $$
$$\mu_{12}=[0,2,0,0,0] +[1,0,0,0,2], \dim(\mu_{12})=1820, \deg(\mu_{12})=16; $$
$$\mu_{13}=[0,1,0,0,1], \dim(\mu_{13})=560, \deg(\mu_{13})=17; $$
$$\mu_{14}=[0,0,1,0,0], \dim(\mu_{14})=120, \deg(\mu_{14})=18; $$
$$\mu_{15}=[0,0,0,1,0], \dim(\mu_{15})=16, \deg(\mu_{15})=19; $$
$$\mu_{16}=[0,0,0,0,0], \dim(\mu_{16})=1, \deg(\mu_{16})=20. $$

\item D=10+0, n=2
$$\mu_0=[0,0,1,0,0], \dim(\mu_0)=120, \deg(\mu_0)=6; $$
$$\mu_1=[0,0,0,1,0] +[0,1,0,1,0] +[1,0,0,0,1], \dim(\mu_1)=720, \deg(\mu_1)=7; $$
\begin{equation*}
\begin{split}
\mu_2=&[0,0,0,0,0] +[0,0,0,1,1] +[0,1,0,0,0] +[0,2,0,0,0] +[1,0,0,2,0]+ \\
& +[2,0,0,0,0], \dim(\mu_2)=2130, \deg(\mu_2)=8;
\end{split}
\end{equation*}
$$\mu_3={\mu_3}'+{\mu_3}'', $$
$${\mu_3}'=[0,0,0,3,0] +[1,0,0,1,0] +[1,1,0,1,0], \dim({\mu_3}')=4512, \deg({\mu_3}')=9; $$
$${\mu_3}''=[0,0,0,0,2] +[1,0,0,0,0], \dim({\mu_3}'')=136, \deg({\mu_3}'')=10; $$
$$\mu_4={\mu_4}'+{\mu_4}'', $$
$${\mu_4}'=[0,1,0,2,0] + [2,0,1,0,0], \dim({\mu_4}')=8008, \deg({\mu_4}')=10; $$
$${\mu_4}''=[0,0,0,1,0] + [1,0,0,0,1], \dim({\mu_4}'')=160, \deg({\mu_4}'')=11; $$
$$\mu_5={\mu_5}'+{\mu_5}'', $$
$${\mu_5}'=[1,0,1,1,0] +[3,0,0,0,1], \dim({\mu_5}')=11440, \deg({\mu_5}')=11; $$
$${\mu_5}''=[0,1,0,0,0], \dim({\mu_5}'')=45, \deg({\mu_5}'')=12; $$
$$\mu_6=[0,0,2,0,0] + [2,0,0,1,1] + [4,0,0,0,0], \dim(\mu_6)=12870, \deg(\mu_6)=12; $$
$$\mu_7=[1,0,1,0,1] +[3,0,0,1,0], \dim(\mu_7)=11440, \deg(\mu_7)=13; $$
$$\mu_8=[0,1,0,0,2] +[2,0,1,0,0], \dim(\mu_8)=8008, \deg(\mu_8)=14; $$
$$\mu_9=[0,0,0,0,3] +[1,1,0,0,1], \dim(\mu_9)=4368, \deg(\mu_9)=15; $$
$$\mu_{10}=[0,2,0,0,0] +[1,0,0,0,2], \dim(\mu_{10})=1820, \deg(\mu_{10})=16; $$
$$\mu_{11}=[0,1,0,0,1], \dim(\mu_{11})=560, \deg(\mu_{11})=17; $$
$$\mu_{12}=[0,0,1,0,0], \dim(\mu_{12})=120, \deg(\mu_{12})=18; $$
$$\mu_{13}=[0,0,0,1,0], \dim(\mu_{13})=16, \deg(\mu_{13})=19; $$
$$\mu_{14}=[0,0,0,0,0], \dim(\mu_{14})=1, \deg(\mu_{14})=20. $$

\item D=10+0, n=3
$$\mu_0=[0,1,0,0,0], \dim(\mu_0)=45, \deg(\mu_0)=8; $$
$$\mu_1=[0,0,0,0,1] +[1,0,0,1,0], \dim(\mu_1)=160, \deg(\mu_1)=9; $$
$$\mu_2={\mu_2}'+{\mu_2}'', $$
$${\mu_2}'=[0,0,0,2,0] +[1,0,0,0,0], \dim({\mu_2}')=136, \deg({\mu_2}')=10; $$
$${\mu_2}''=[1,0,0,0,1], \dim({\mu_2}'')=144, \deg({\mu_2}'')=11; $$
$$\mu_3=[0,0,0,0,0] +[0,0,0,1,1] +[0,1,0,0,0] +[2,0,0,0,0], \dim(\mu_3)=310, \deg(\mu_3)=12; $$
$$\mu_4=[0,0,0,0,1] +[1,0,0,1,0], \dim(\mu_4)=160, \deg(\mu_4)=13; $$
$$\mu_5=[0,0,0,1,0], \dim(\mu_5)=16, \deg(\mu_5)=15; $$
$$\mu_6=[0,0,0,0,0], \dim(\mu_6)=1, \deg(\mu_6)=16. $$

\item D=10+0, n=4
$$\mu_0=[1,0,0,0,0], \dim(\mu_0)=10, \deg(\mu_0)=10; $$
$$\mu_1={\mu_1}'+{\mu_1}'', $$
$${\mu_1}'=[0,0,0,1,0], \dim({\mu_1}')=16, \deg({\mu_1}')=11; $$
$${\mu_1}''=[2,0,0,0,0], \dim({\mu_1}'')=54, \deg({\mu_1}'')=12; $$
$$\mu_2=[0,0,0,0,1] +[1,0,0,1,0], \dim(\mu_2)=160, \deg(\mu_2)=13; $$
$$\mu_3=[0,0,1,0,0] +[1,0,0,0,0], \dim(\mu_3)=130, \deg(\mu_3)=14; $$
$$\mu_4=[0,0,0,0,0] +[0,1,0,0,0], \dim(\mu_4)=46, \deg(\mu_4)=16; $$
$$\mu_5=[0,0,0,0,1], \dim(\mu_5)=16, \deg(\mu_5)=17. $$

\item D=10+0, n=5
$$\mu_0=[0,0,0,0,0], \dim(\mu_0)=1, \deg(\mu_0)=12; $$
$$\mu_1=[1,0,0,0,0], \dim(\mu_1)=10, \deg(\mu_1)=14; $$
$$\mu_2=[0,0,0,1,0], \dim(\mu_2)=16, \deg(\mu_2)=15; $$
$$\mu_3=[0,0,0,0,1], \dim(\mu_3)=16, \deg(\mu_3)=17; $$
$$\mu_4=[1,0,0,0,0], \dim(\mu_4)=10, \deg(\mu_4)=18; $$
$$\mu_5=[0,0,0,0,0], \dim(\mu_5)=1, \deg(\mu_5)=20. $$

\item D=9+1, n=0
$$\mu_0=[0,0,0,0], \dim(\mu_0)=1, \deg(\mu_0)=0; $$
$$\mu_1=[1,0,0,0], \dim(\mu_1)=9, \deg(\mu_1)=2; $$
$$\mu_2=[0,0,1,0] +[0,1,0,0], \dim(\mu_2)=120, \deg(\mu_2)=4; $$
$$\mu_3=[0,0,0,1] +[0,1,0,1] +[1,0,0,1], \dim(\mu_3)=576, \deg(\mu_3)=5; $$
\begin{equation*}
\begin{split}
\mu_4=&[0,0,0,0] +[0,0,0,2] +[0,2,0,0] +[1,0,0,0] +[1,0,0,2] +[1,1,0,0]+ \\
& +[2,0,0,0], \dim(\mu_4)=1830, \deg(\mu_4)=6;
\end{split}
\end{equation*}
$$\mu_5={\mu_5}'+{\mu_5}'', $$
$${\mu_5}'=[0,0,0,3] +[0,1,0,1] +[1,0,0,1] +[1,1,0,1] +[2,0,0,1], \dim({\mu_5}')=4368, \deg({\mu_5}')=7; $$
$${\mu_5}''=[0,0,0,0], \dim({\mu_5}'')=1, \deg({\mu_5}'')=8; $$
\begin{equation*}
\begin{split}
\mu_6=&[0,0,1,0] + [0,1,0,0] + [0,1,0,2] + [1,0,0,2] + [1,0,1,0] + [1,1,0,0]+\\
&+ [2,0,1,0] + [2,1,0,0], \dim(\mu_6)=8008, \deg(\mu_6)=8;
\end{split}
\end{equation*}
\begin{equation*}
\begin{split}
\mu_7=&[0,0,0,1] +[0,0,1,1] +[0,1,0,1] +[1,0,0,1] +[1,0,1,1] +[1,1,0,1]+\\
&+[2,0,0,1] +[3,0,0,1], \dim(\mu_7)=11440, \deg(\mu_7)=9;
\end{split}
\end{equation*}
\begin{equation*}
\begin{split}
\mu_8=&[0,0,0,0] +[0,0,0,2] +[0,0,1,0] +[0,0,2,0] +[0,1,1,0] +[0,2,0,0]+\\
&+[1,0,0,0] +[1,0,0,2] +[1,0,1,0] +[2,0,0,0] +[2,0,0,2] +[2,0,1,0]+\\
&+[3,0,0,0] +[4,0,0,0], \dim(\mu_8)=12870, \deg(\mu_8)=10;
\end{split}
\end{equation*}
\begin{equation*}
\begin{split}
\mu_9=&[0,0,0,1] +[0,0,1,1] +[0,1,0,1] +[1,0,0,1] +[1,0,1,1] +[1,1,0,1]+\\
&+[2,0,0,1] +[3,0,0,1], \dim(\mu_9)=11440, \deg(\mu_9)=11;
\end{split}
\end{equation*}
\begin{equation*}
\begin{split}
\mu_{10}=&[0,0,1,0] +[0,1,0,0] +[0,1,0,2] +[1,0,0,2] +[1,0,1,0] +[1,1,0,0]+\\
&+[2,0,1,0] +[2,1,0,0], \dim(\mu_{10})=8008, \deg(\mu_{10})=12;
\end{split}
\end{equation*}
$$\mu_{11}=[0,0,0,3] +[0,1,0,1] +[1,0,0,1] +[1,1,0,1] +[2,0,0,1], \dim(\mu_{11})=4368, \deg(\mu_{11})=13; $$
$$\mu_{12}=[0,0,0,2] +[0,2,0,0] +[1,0,0,2] +[1,1,0,0] +[2,0,0,0], \dim(\mu_{12})=1820, \deg(\mu_{12})=14; $$
$$\mu_{13}=[0,1,0,1] +[1,0,0,1], \dim(\mu_{13})=560, \deg(\mu_{13})=15; $$
$$\mu_{14}=[0,0,1,0] +[0,1,0,0], \dim(\mu_{14})=120, \deg(\mu_{14})=16; $$
$$\mu_{15}=[0,0,0,1], \dim(\mu_{15})=16, \deg(\mu_{15})=17; $$
$$\mu_{16}=[0,0,0,0], \dim(\mu_{16})=1, \deg(\mu_{16})=18. $$

\item D=9+1, n=1
$$\mu_0=[0,0,1,0], \dim(\mu_0)=84, \deg(\mu_0)=4; $$
$$\mu_1=[0,0,0,1] +[0,1,0,1] +[1,0,0,1], \dim(\mu_1)=576, \deg(\mu_1)=5; $$
\begin{equation*}
\begin{split}
\mu_2=&[0,0,0,0] +[0,0,0,2] +[0,0,1,0] +[0,1,0,0] +[0,2,0,0] +[1,0,0,0]+\\
&+[1,0,0,2] +[1,1,0,0] +[2,0,0,0], \dim(\mu_2)=1950, \deg(\mu_2)=6;
\end{split}
\end{equation*}
\begin{equation*}
\begin{split}
\mu_3=&[0,0,0,1] +[0,0,0,3] +[0,1,0,1] +2\times[1,0,0,1] +[1,1,0,1] +[2,0,0,1],\\
& \dim(\mu_3)=4512, \deg(\mu_3)=7;
\end{split}
\end{equation*}
$$\mu_4={\mu_4}'+{\mu_4}'', $$
\begin{equation*}
\begin{split}
{\mu_4}'=&[0,0,1,0] +[0,1,0,0] +[0,1,0,2] +[1,0,0,2] +[1,0,1,0] +[1,1,0,0]+\\
&+[2,0,0,0] +[2,0,1,0] +[2,1,0,0], \dim({\mu_4}')=8052, \deg({\mu_4}')=8;
\end{split}
\end{equation*}
$${\mu_4}''=[0,0,0,1], \dim({\mu_4}'')=16, \deg({\mu_4}'')=9; $$
$$\mu_5={\mu_5}'+{\mu_5}'', $$
\begin{equation*}
\begin{split}
{\mu_5}'=&[0,0,0,1] +[0,0,1,1] +[0,1,0,1] +[1,0,0,1] +[1,0,1,1] +[1,1,0,1]+\\
&+[2,0,0,1] +[3,0,0,1], \dim({\mu_5}')=11440, \deg({\mu_5}')=9;
\end{split}
\end{equation*}
$${\mu_5}''=[1,0,0,0], \dim({\mu_5}'')=9, \deg({\mu_5}'')=10; $$
\begin{equation*}
\begin{split}
\mu_6=&[0,0,0,0] +[0,0,0,2] +[0,0,1,0] +[0,0,2,0] +[0,1,1,0] +[0,2,0,0]+\\
&+[1,0,0,0]+[1,0,0,2] +[1,0,1,0] +[2,0,0,0] +[2,0,0,2] +[2,0,1,0] +\\
&+[3,0,0,0]+[4,0,0,0], \dim(\mu_6)=12870, \deg(\mu_6)=10;
\end{split}
\end{equation*}
\begin{equation*}
\begin{split}
\mu_7=&[0,0,0,1] +[0,0,1,1] +[0,1,0,1] +[1,0,0,1] +[1,0,1,1] +[1,1,0,1]+\\
& +[2,0,0,1] +[3,0,0,1], \dim(\mu_7)=11440, \deg(\mu_7)=11;
\end{split}
\end{equation*}
\begin{equation*}
\begin{split}
\mu_8=&[0,0,1,0] +[0,1,0,0] +[0,1,0,2] +[1,0,0,2] +[1,0,1,0] +[1,1,0,0]+\\
& +[2,0,1,0] +[2,1,0,0], \dim(\mu_8)=8008, \deg(\mu_8)=12;
\end{split}
\end{equation*}
$$\mu_9=[0,0,0,3] +[0,1,0,1] +[1,0,0,1] +[1,1,0,1] +[2,0,0,1], \dim(\mu_9)=4368, \deg(\mu_9)=13; $$
$$\mu_{10}= [0,0,0,2] +[0,2,0,0] +[1,0,0,2] +[1,1,0,0] +[2,0,0,0], \dim(\mu_{10})=1820, \deg(\mu_{10})=14; $$
$$\mu_{11}= +[0,1,0,1] +[1,0,0,1], \dim(\mu_{11})=560, \deg(\mu_{11})=15; $$
$$\mu_{12}=[0,0,1,0] +[0,1,0,0], \dim(\mu_{12})=120, \deg(\mu_{12})=16; $$
$$\mu_{13}=[0,0,0,1], \dim(\mu_{13})=16, \deg(\mu_{13})=17; $$
$$\mu_{14}=[0,0,0,0], \dim(\mu_{14})=1, \deg(\mu_{14})=18. $$

\item D=9+1, n=2
$$\mu_0=[0,1,0,0], \dim(\mu_0)=36, \deg(\mu_0)=6; $$
$$\mu_1=[0,0,0,1] +[1,0,0,1], \dim(\mu_1)=144, \deg(\mu_1)=7; $$
$$\mu_2=[0,0,0,0] +[0,0,0,2] +[1,0,0,0] +[2,0,0,0], \dim(\mu_2)=180, \deg(\mu_2)=8; $$
$$\mu_3=[0,0,0,0] +[0,0,0,2] +[1,0,0,0] +[2,0,0,0], \dim(\mu_3)=180, \deg(\mu_3)=10; $$
$$\mu_4=[0,0,0,1] +[1,0,0,1], \dim(\mu_4)=144, \deg(\mu_4)=11; $$
$$\mu_5=[0,1,0,0], \dim(\mu_5)=36, \deg(\mu_5)=12. $$

\item D=9+1, n=3
$$\mu_0=[1,0,0,0], \dim(\mu_0)=9, \deg(\mu_0)=8; $$
$$\mu_1={\mu_1}'+{\mu_1}'', $$
$${\mu_1}'=[0,0,0,1], \dim({\mu_1}')=16, \deg({\mu_1}')=9; $$
$${\mu_1}''=[2,0,0,0], \dim({\mu_1}'')=44, \deg({\mu_1}'')=10; $$
$$\mu_2=[0,0,0,1] +[1,0,0,1], \dim(\mu_2)=144, \deg(\mu_2)=11; $$
$$\mu_3=[0,0,0,0] +[0,0,1,0] +[0,1,0,0] +[1,0,0,0], \dim(\mu_3)=130, \deg(\mu_3)=12; $$
$$\mu_4={\mu_4}'+{\mu_4}'', $$
$${\mu_4}'=[0,0,0,1], \dim({\mu_4}')=16, \deg({\mu_4}')=13; $$
$${\mu_4}''=[0,1,0,0], \dim({\mu_4}'')=36, \deg({\mu_4}'')=14; $$
$$\mu_5=[0,0,0,1], \dim(\mu_5)=16, \deg(\mu_5)=15; $$
$$\mu_6=[0,0,0,0], \dim(\mu_6)=1, \deg(\mu_6)=16. $$

\item D=9+1, n=4
$$\mu_0=[0,0,0,0], \dim(\mu_0)=1, \deg(\mu_0)=10; $$
$$\mu_1=[0,0,0,0] +[1,0,0,0], \dim(\mu_1)=10, \deg(\mu_1)=12; $$
$$\mu_2=[0,0,0,1], \dim(\mu_2)=16, \deg(\mu_2)=13; $$
$$\mu_3=[0,0,0,1], \dim(\mu_3)=16, \deg(\mu_3)=15; $$
$$\mu_4=[0,0,0,0] +[1,0,0,0], \dim(\mu_4)=10, \deg(\mu_4)=16; $$
$$\mu_5=[0,0,0,0], \dim(\mu_5)=1, \deg(\mu_5)=18. $$

\item D=8+2, n=0
$$\mu_0=[0,0,0,0,0], \dim(\mu_0)=1, \deg(\mu_0)=0; $$
$$\mu_1=[1,0,0,0,0], \dim(\mu_1)=8, \deg(\mu_1)=2; $$
$$\mu_2=[0,1,0,0,0], \dim(\mu_2)=56, \deg(\mu_2)=4; $$
$$\mu_3=[0,0,0,1,-1] +[0,0,1,0, 1] +[1,0,0,1, 1] +[1,0,1,0,-1], \dim(\mu_3)=128, \deg(\mu_3)=5; $$
\begin{equation*}
\begin{split}
\mu_4=&[0,0,0,0,-2] +[0,0,0,0, 2] +[0,0,0,2, 2] +[0,0,2,0,-2] +[1,0,0,0,0]+\\
& +[2,0,0,0,-2] +[2,0,0,0, 2], \dim(\mu_4)=150, \deg(\mu_4)=6;
\end{split}
\end{equation*}
$$\mu_5={\mu_5}'+{\mu_5}'', $$
$${\mu_5}'=[1,0,0,1, 3] +[1,0,1,0,-3], \dim({\mu_5}')=112, \deg({\mu_5}')=7; $$
$${\mu_5}''=[0,0,0,0, 0], \dim({\mu_5}'')=1, \deg({\mu_5}'')=8; $$
$$\mu_6=[0,1,0,0,-4] + [0,1,0,0, 4], \dim(\mu_6)=56, \deg(\mu_6)=8; $$
$$\mu_7=[0,0,0,1,-5] +[0,0,1,0, 5], \dim(\mu_7)=16, \deg(\mu_7)=9; $$
$$\mu_8=[0,0,0,0,-6] +[0,0,0,0, 6], \dim(\mu_8)=2, \deg(\mu_8)=10. $$

\item D=8+2, n=1
$$\mu_0=[0,1,0,0,0], \dim(\mu_0)=28, \deg(\mu_0)=4; $$
$$\mu_1=[0,0,0,1,-1] +[0,0,1,0, 1] +[1,0,0,1, 1] +[1,0,1,0,-1], \dim(\mu_1)=128, \deg(\mu_1)=5; $$
\begin{equation*}
\begin{split}
\mu_2=&[0,0,0,0,-2] +[0,0,0,0, 2] +[0,0,0,2, 2] +[0,0,1,1, 0] +[0,0,2,0,-2]+\\
& +2\times[1,0,0,0, 0] +[2,0,0,0,-2] +[2,0,0,0, 2], \dim(\mu_2)=214, \deg(\mu_2)=6;
\end{split}
\end{equation*}
$$\mu_3={\mu_3}'+{\mu_3}'', $$
$${\mu_3}'=[0,0,0,1, 1] +[0,0,1,0,-1] +[1,0,0,1, 3] +[1,0,1,0,-3], \dim({\mu_3}')=128, \deg({\mu_3}')=7; $$
$${\mu_3}''=[0,0,0,0, 0] +[0,0,0,2, 0] +[0,0,2,0, 0] +[2,0,0,0, 0], \dim({\mu_3}'')=106, \deg({\mu_3}'')=8; $$
$$\mu_4={\mu_4}'+{\mu_4}'', $$
$${\mu_4}'=[0,1,0,0,-4] +[0,1,0,0, 4] , \dim({\mu_4}')=56, \deg({\mu_4}')=8; $$
$${\mu_4}''=[0,0,0,1,-1] + [0,0,1,0, 1] +[1,0,0,1, 1] + [1,0,1,0,-1], \dim({\mu_4}'')=128, \deg({\mu_4}'')=9; $$
$$\mu_5={\mu_5}'+{\mu_5}'', $$
$${\mu_5}'=[0,0,0,1,-5]+[0,0,1,0, 5], \dim({\mu_5}')=16, \deg({\mu_5}')=9; $$
$${\mu_5}''=[0,1,0,0,-2] +[0,1,0,0, 2] +[1,0,0,0, 0], \dim({\mu_5}'')=64, \deg({\mu_5}'')=10; $$
$$\mu_6={\mu_6}'+{\mu_6}'', $$
$${\mu_6}'=[0,0,0,0,-6] + [0,0,0,0, 6], \dim({\mu_6}')=2, \deg({\mu_6}')=10; $$
$${\mu_6}''=[0,0,0,1,-3] +[0,0,1,0, 3], \dim({\mu_6}'')=16, \deg({\mu_6}'')=11; $$
$$\mu_7=[0,0,0,0,-4] +[0,0,0,0, 4], \dim(\mu_7)=2, \deg(\mu_7)=12. $$

\item D=8+2, n=2
$$\mu_0=[1,0,0,0,0], \dim(\mu_0)=8, \deg(\mu_0)=6; $$
$$\mu_1={\mu_1}'+{\mu_1}'', $$
$${\mu_1}'=[0,0,0,1, 1] +[0,0,1,0,-1], \dim({\mu_1}')=16, \deg({\mu_1}')=7; $$
$${\mu_1}''=[2,0,0,0,0], \dim({\mu_1}'')=35, \deg({\mu_1}'')=8; $$
$$\mu_2={\mu_2}'+{\mu_2}'', $$
$${\mu_2}'=[0,0,0,0,0], \dim({\mu_2}')=1, \deg({\mu_2}')=8; $$
$${\mu_2}''=[0,0,0,1,-1] +[0,0,1,0, 1] +[1,0,0,1, 1] +[1,0,1,0,-1], \dim({\mu_2}'')=128, \deg({\mu_2}'')=9; $$
\begin{equation*}
\begin{split}
\mu_3=&[0,0,0,0,-2] +[0,0,0,0, 2] +[0,0,1,1, 0] +[0,1,0,0,-2] +[0,1,0,0, 2]+\\
& +2\times[1,0,0,0, 0], \dim(\mu_3)=130, \deg(\mu_3)=10;
\end{split}
\end{equation*}
$$\mu_4={\mu_4}'+{\mu_4}'', $$
$${\mu_4}'=[0,0,0,1,-3] +[0,0,0,1, 1] +[0,0,1,0,-1] +[0,0,1,0, 3], \dim({\mu_4}')=32, \deg({\mu_4}')=11; $$
$${\mu_4}''=[0,1,0,0, 0], \dim({\mu_4}'')=28, \deg({\mu_4}'')=12; $$
$$\mu_5={\mu_5}'+{\mu_5}'', $$
$${\mu_5}'=[0,0,0,0,-4] +[0,0,0,0, 4], \dim({\mu_5}')=2, \deg({\mu_5}')=12; $$
$${\mu_5}''=[0,0,0,1,-1] +[0,0,1,0, 1], \dim({\mu_5}'')=16, \deg({\mu_5}'')=13; $$
$$\mu_6=[0,0,0,0,-2] +[0,0,0,0, 2], \dim(\mu_6)=2, \deg(\mu_6)=14. $$

\item D=8+2, n=3
$$\mu_0=[0,0,0,0,0], \dim(\mu_0)=1, \deg(\mu_0)=8; $$
$$\mu_1=[0,0,0,0,-2] +[0,0,0,0, 2] +[1,0,0,0, 0], \dim(\mu_1)=10, \deg(\mu_1)=10; $$
$$\mu_2=[0,0,0,1, 1] +[0,0,1,0,-1], \dim(\mu_2)=16, \deg(\mu_2)=11; $$
$$\mu_3=[0,0,0,1,-1] +[0,0,1,0, 1], \dim(\mu_3)=16, \deg(\mu_3)=13; $$
$$\mu_4=[0,0,0,0,-2] +[0,0,0,0, 2] +[1,0,0,0, 0], \dim(\mu_4)=10, \deg(\mu_4)=14; $$
$$\mu_5=[0,0,0,0,0], \dim(\mu_5)=1, \deg(\mu_5)=16. $$

\item D=7+3, n=0
$$\mu_0=[0,0,0,0], \dim(\mu_0)=1, \deg(\mu_0)=0; $$
$$\mu_1=[1,0,0,0], \dim(\mu_1)=7, \deg(\mu_1)=2; $$
$$\mu_2=[0,1,0,0] +[1,0,0,0], \dim(\mu_2)=28, \deg(\mu_2)=4; $$
$$\mu_3={\mu_3}'+{\mu_3}'', $$
$${\mu_3}'=[0,0,1,1], \dim({\mu_3}')=16, \deg({\mu_3}')=5; $$
$${\mu_3}''=[0,0,0,0]+[0,0,2,0] +[2,0,0,0], \dim({\mu_3}'')=63, \deg({\mu_3}'')=6; $$
$$\mu_4={\mu_4}'+{\mu_4}'', $$
$${\mu_4}'=[0,0,0,2], \dim({\mu_4}')=3, \deg({\mu_4}')=6; $$
$${\mu_4}''=[0,0,1,1] +[1,0,1,1], \dim({\mu_4}'')=112, \deg({\mu_4}'')=7; $$
$$\mu_5=[0,0,0,0] +[0,1,0,2] +[1,0,0,2], \dim(\mu_5)=85, \deg(\mu_5)=8; $$
$$\mu_6=[0,0,1,3], \dim(\mu_6)=32, \deg(\mu_6)=9; $$
$$\mu_7=[0,0,0,4], \dim(\mu_7)=5, \deg(\mu_7)=10. $$

\item D=7+3, n=1
$$\mu_0=[1,0,0,0], \dim(\mu_0)=7, \deg(\mu_0)=4; $$
$$\mu_1={\mu_1}'+{\mu_1}'', $$
$${\mu_1}'=[0,0,1,1], \dim({\mu_1}')=16, \deg({\mu_1}')=5; $$
$${\mu_1}''=[2,0,0,0], \dim({\mu_1}'')=27, \deg({\mu_1}'')=6; $$
$$\mu_2={\mu_2}'+{\mu_2}'', $$
$${\mu_2}'=[0,0,0,2], \dim({\mu_2}')=3, \deg({\mu_2}')=6; $$
$${\mu_2}''=[0,0,1,1] +[1,0,1,1], \dim({\mu_2}'')=112, \deg({\mu_2}'')=7; $$
\begin{equation*}
\begin{split}
\mu_3=&[0,0,0,0] +[0,0,0,2] +[0,0,2,0] +[0,1,0,2] +[1,0,0,0] +[1,0,0,2],\\
& \dim(\mu_3)=130, \deg(\mu_3)=8;
\end{split}
\end{equation*}
$$\mu_4={\mu_4}'+{\mu_4}'', $$
$${\mu_4}'=[0,0,1,1] +[0,0,1,3], \dim({\mu_4}')=48, \deg({\mu_4}')=9; $$
$${\mu_4}''=[0,1,0,0], \dim({\mu_4}'')=21, \deg({\mu_4}'')=10; $$
$$\mu_5={\mu_5}'+{\mu_5}'', $$
$${\mu_5}'=[0,0,0,4], \dim({\mu_5}')=5, \deg({\mu_5}')=10; $$
$${\mu_5}''=[0,0,1,1], \dim({\mu_5}'')=16, \deg({\mu_5}'')=11; $$
$$\mu_6=[0,0,0,2], \dim(\mu_6)=3, \deg(\mu_6)=12. $$

\item D=7+3, n=2
$$\mu_0=[0,0,0,0], \dim(\mu_0)=1, \deg(\mu_0)=6; $$
$$\mu_1=[0,0,0,2] +[1,0,0,0], \dim(\mu_1)=10, \deg(\mu_1)=8; $$
$$\mu_2=[0,0,1,1], \dim(\mu_2)=16, \deg(\mu_2)=9; $$
$$\mu_3=[0,0,1,1], \dim(\mu_3)=16, \deg(\mu_3)=11; $$
$$\mu_4=[0,0,0,2] +[1,0,0,0], \dim(\mu_4)=10, \deg(\mu_4)=12; $$
$$\mu_5=[0,0,0,0], \dim(\mu_5)=1, \deg(\mu_5)=14. $$

\item D=6+4, n=0
$$\mu_0=[0,0,0,0,0], \dim(\mu_0)=1, \deg(\mu_0)=0; $$
$$\mu_1=[0,1,0,0,0], \dim(\mu_1)=6, \deg(\mu_1)=2; $$
$$\mu_2=[0,0,0,0,0] +[1,0,1,0,0], \dim(\mu_2)=16, \deg(\mu_2)=4; $$
$$\mu_3=[0,0,0,1,1] +[0,0,2,0,0] +[0,1,0,0,0] +[2,0,0,0,0], \dim(\mu_3)=30, \deg(\mu_3)=6; $$
$$\mu_4={\mu_4}'+{\mu_4}'', $$
$${\mu_4}'=[0,0,1,1,0] +[1,0,0,0,1], \dim({\mu_4}')=16, \deg({\mu_4}')=7; $$
$${\mu_4}''=[1,0,1,0,0], \dim({\mu_4}'')=15, \deg({\mu_4}'')=8; $$
$$\mu_5=[0,0,1,0,1] +[1,0,0,1,0], \dim(\mu_5)=16, \deg(\mu_5)=9; $$
$$\mu_6=[0,0,0,1,1], \dim(\mu_6)=4, \deg(\mu_6)=10. $$

\item D=6+4, n=1
$$\mu_0=[0,0,0,0,0], \dim(\mu_0)=1, \deg(\mu_0)=4; $$
$$\mu_1=[0,0,0,1,1] +[0,1,0,0,0], \dim(\mu_1)=10, \deg(\mu_1)=6; $$
$$\mu_2=[0,0,1,1,0] +[1,0,0,0,1], \dim(\mu_2)=16, \deg(\mu_2)=7; $$
$$\mu_3=[0,0,1,0,1] +[1,0,0,1,0], \dim(\mu_3)=16, \deg(\mu_3)=9; $$
$$\mu_4=[0,0,0,1,1] +[0,1,0,0,0], \dim(\mu_4)=10, \deg(\mu_4)=10; $$
$$\mu_5=[0,0,0,0,0], \dim(\mu_5)=1, \deg(\mu_5)=12. $$

\item D=5+5, n=0
$$\mu_0=[0,0,0,0], \dim(\mu_0)=1, \deg(\mu_0)=0; $$
$$\mu_1=[1,0,0,0], \dim(\mu_1)=5, \deg(\mu_1)=2; $$
$$\mu_2=[0,2,0,0], \dim(\mu_2)=10, \deg(\mu_2)=4; $$
$$\mu_3=[0,2,0,0], \dim(\mu_3)=10, \deg(\mu_3)=6; $$
$$\mu_4=[1,0,0,0], \dim(\mu_4)=5, \deg(\mu_4)=8; $$
$$\mu_5=[0,0,0,0], \dim(\mu_5)=1, \deg(\mu_5)=10. $$

\item D=4+6, n=0
$$\mu_0=[0,0,0,0,0], \dim(\mu_0)=1, \deg(\mu_0)=0; $$
$$\mu_1=[1,1,0,0,0], \dim(\mu_1)=4, \deg(\mu_1)=2; $$
$$\mu_2=[0,2,0,0,0] +[2,0,0,0,0], \dim(\mu_2)=6, \deg(\mu_2)=4; $$
$$\mu_3=[1,1,0,0,0], \dim(\mu_3)=4, \deg(\mu_3)=6; $$
$$\mu_4=[0,0,0,0,0], \dim(\mu_4)=1, \deg(\mu_4)=8. $$

\item D=3+7, n=0
$$\mu_0=[0,0,0,0], \dim(\mu_0)=1, \deg(\mu_0)=0; $$
$$\mu_1=[2,0,0,0], \dim(\mu_1)=3, \deg(\mu_1)=2; $$
$$\mu_2=[2,0,0,0], \dim(\mu_2)=3, \deg(\mu_2)=4; $$
$$\mu_3=[0,0,0,0], \dim(\mu_3)=1, \deg(\mu_3)=6. $$

\item D=2+8, n=0
$$\mu_0=[0,0,0,0,0], \dim(\mu_0)=1, \deg(\mu_0)=0; $$
$$\mu_1=[0,0,0,0,-2] +[0,0,0,0, 2], \dim(\mu_1)=2, \deg(\mu_1)=2; $$
$$\mu_2=[0,0,0,0,0], \dim(\mu_2)=1, \deg(\mu_2)=4. $$

\item D=1+9, n=0
$$\mu_0=[0,0,0,0], \dim(\mu_0)=1, \deg(\mu_0)=0; $$
$$\mu_1=[0,0,0,0], \dim(\mu_1)=1, \deg(\mu_1)=2. $$

\end{itemize}

\subsection{Resolution of the cohomology modules of 6D Lie algebra of supersymmetries}
We now give the structure of $\mu_i$ as $aut$-module and its grading in the case of six-dimensional Lie algebra of supersymmetries.

\begin{itemize}
\item D=6+0, n=0
$$\mu_0=[0,0,0,0], \dim(\mu_0)=1, \deg(\mu_0)=0; $$
$$\mu_1=[0,1,0,0], \dim(\mu_1)=6, \deg(\mu_1)=2; $$
$$\mu_2=[1,0,0,1], \dim(\mu_2)=8, \deg(\mu_2)=3; $$
$$\mu_3=[0,0,0,2], \dim(\mu_3)=3, \deg(\mu_3)=4. $$

\item D=6+0, n=1
$$\mu_0=[1,0,0,1], \dim(\mu_0)=8, \deg(\mu_0)=3; $$
$$\mu_1=[0,0,0,2] +[1,0,1,0], \dim(\mu_1)=18, \deg(\mu_1)=4; $$
$$\mu_2=[0,0,2,0] +[0,1,0,2] +[2,0,0,0], \dim(\mu_2)=38, \deg(\mu_2)=6; $$
$$\mu_3=[0,1,1,1] +[1,0,0,1] +[1,0,0,3], \dim(\mu_3)=64, \deg(\mu_3)=7; $$
$$\mu_4=[0,0,0,0] +[0,0,0,4] +[0,2,0,0] +[1,0,1,2], \dim(\mu_4)=71, \deg({\mu_4}')=8; $$
$$\mu_5=[0,0,1,3] +[1,1,0,1], \dim(\mu_5)=56, \deg(\mu_5)=9; $$
$$\mu_6=[0,1,0,2] +[2,0,0,0], \dim(\mu_6)=28, \deg(\mu_6)=10; $$
$$\mu_7=[1,0,0,1], \dim(\mu_7)=8, \deg(\mu_7)=11; $$
$$\mu_8=[0,0,0,0], \dim(\mu_8)=1, \deg(\mu_8)=12. $$

\item D=6+0, n=2
$$\mu_0=[0,1,0,2], \dim(\mu_0)=18, \deg(\mu_0)=6; $$
$$\mu_1=[0,1,1,1] +[1,0,0,1] +[1,0,0,3], \dim(\mu_1)=64, \deg(\mu_1)=7; $$
$$\mu_2=[0,0,0,0] +[0,0,0,2] +[0,0,0,4] +[0,2,0,0] +[1,0,1,0] +[1,0,1,2], \dim(\mu_2)=89, \deg(\mu_2)=8; $$
$$\mu_3=[0,0,1,1] +[0,0,1,3] +[1,1,0,1], \dim(\mu_3)=64, \deg(\mu_3)=9; $$
$$\mu_4=[0,1,0,2] +[2,0,0,0], \dim(\mu_4)=28, \deg({\mu_4}')=10; $$
$$\mu_5=[1,0,0,1], \dim(\mu_5)=8, \deg(\mu_5)=11; $$
$$\mu_6=[0,0,0,0], \dim(\mu_6)=1, \deg(\mu_6)=12. $$

\item D=6+0, n=3
$$\mu_0=[0,0,0,2], \dim(\mu_0)=3, \deg(\mu_0)=8; $$
$$\mu_1=[0,0,1,1], \dim(\mu_1)=8, \deg(\mu_1)=9; $$
$$\mu_2=[0,1,0,0], \dim(\mu_2)=6, \deg(\mu_2)=10; $$
$$\mu_3=[0,0,0,0], \dim(\mu_3)=1, \deg(\mu_3)=12. $$

\end{itemize}

\subsection{Resolution of the cohomology modules of 5D Lie algebra of supersymmetries}
We now give the structure of $\mu_i$ as $aut$-module and its grading in the case of five-dimensional Lie algebra of supersymmetries.
\begin{itemize}
\item D=5+0, n=0
$$\mu_0=[0,0,0], \dim(\mu_0)=1, \deg(\mu_0)=0; $$
$$\mu_1=[1,0,0], \dim(\mu_1)=5, \deg(\mu_1)=2; $$
$$\mu_2=[0,2,0] +[1,0,2], \dim(\mu_2)=25, \deg(\mu_2)=4; $$
$$\mu_3=[0,1,1] +[0,1,3] +[1,1,1], \dim(\mu_3)=56, \deg(\mu_3)=5; $$
$$\mu_4=[0,0,0] +[0,0,4] +[0,2,2] +[1,0,0] +[1,0,2] +[2,0,0], \dim(\mu_4)=70, \deg({\mu_4}')=6; $$
$$\mu_5=[0,1,1] +[0,1,3] +[1,1,1], \dim(\mu_5)=56, \deg(\mu_5)=7; $$
$$\mu_6=[0,0,2] +[0,2,0] +[1,0,2], \dim(\mu_6)=28, \deg(\mu_6)=8; $$
$$\mu_7=[0,1,1], \dim(\mu_7)=8, \deg(\mu_7)=9; $$
$$\mu_8=[0,0,0], \dim(\mu_8)=1, \deg(\mu_8)=10. $$

\item D=5+0, n=1
$$\mu_0=[1,0,2], \dim(\mu_0)=15, \deg(\mu_0)=4; $$
$$\mu_1=[0,1,1] +[0,1,3] +[1,1,1], \dim(\mu_1)=56, \deg(\mu_1)=5; $$
$$\mu_2=[0,0,0] +[0,0,2] +[0,0,4] +[0,2,0] +[0,2,2] +[1,0,0] +[1,0,2] +[2,0,0], \dim(\mu_2)=83, \deg(\mu_2)=6; $$
$$\mu_3=2\times[0,1,1] +[0,1,3] +[1,1,1], \dim(\mu_3)=64, \deg(\mu_3)=7; $$
$$\mu_4=[0,0,0] +[0,0,2] +[0,2,0] +[1,0,2], \dim(\mu_4)=29, \deg({\mu_4}')=8; $$
$$\mu_5=[0,1,1], \dim(\mu_5)=8, \deg(\mu_5)=9; $$
$$\mu_6=[0,0,0], \dim(\mu_6)=1, \deg(\mu_6)=10. $$

\item D=5+0, n=2
$$\mu_0=[0,0,2], \dim(\mu_0)=3, \deg(\mu_0)=6; $$
$$\mu_1=[0,1,1], \dim(\mu_1)=8, \deg(\mu_1)=7; $$
$$\mu_2=[0,0,0] +[1,0,0], \dim(\mu_2)=6, \deg(\mu_2)=8; $$
$$\mu_3=[0,0,0], \dim(\mu_3)=1, \deg(\mu_3)=10. $$

\end{itemize}

\subsection{Resolution of the cohomology modules of 4D Lie algebra of supersymmetries}
We now give the structure of $\mu_i$ as $aut$-module and its grading in the case of four-dimensional Lie algebra of supersymmetries.
\begin{itemize}
\item D=4+0, n=0
$$\mu_0=[0,0], \dim(\mu_0)=1, \deg(\mu_0)=0; $$
$$\mu_1=[1,1], \dim(\mu_1)=4, \deg(\mu_1)=2; $$
$$\mu_2=[0,1] +[1,0], \dim(\mu_2)=4, \deg(\mu_2)=3; $$
$$\mu_3=[0,0], \dim(\mu_3)=1, \deg(\mu_3)=4. $$

\item D=4+0, n=1
$$\mu_0=[0,1] +[1,0], \dim(\mu_0)=15, \deg(\mu_0)=4; $$
$$\mu_1=[0,0] +[0,2] +[2,0], \dim(\mu_1)=56, \deg(\mu_1)=5; $$
$$\mu_2=2\times[0,0] +[1,1], \dim(\mu_2)=83, \deg(\mu_2)=6; $$
$$\mu_3=[0,1] +[1,0], \dim(\mu_3)=64, \deg(\mu_3)=7; $$
$$\mu_4=[0,0], \dim(\mu_4)=29, \deg({\mu_4}')=8. $$

\item D=4+0, n=2
$$\mu_0=2\times[0,0], \dim(\mu_0)=3, \deg(\mu_0)=6; $$
$$\mu_1=[0,1] +[1,0], \dim(\mu_1)=8, \deg(\mu_1)=7; $$
$$\mu_2=2\times[0,0], \dim(\mu_2)=6, \deg(\mu_2)=8. $$

\end{itemize}

\section{Computer calculations\label{Appendix:code}}
We will describe here the computer programs used in the calculations.{\footnote{The detailed codes are provided here: \url{http://lifshitz.ucdavis.edu/~rxu/code/cohom/}}}

\begin{enumerate}
\item We calculate the differential $d:V\to S\otimes S$ (Eq.~\ref{sud})  using Gamma~\cite{GAMMA}.
\item We use Macaulay2~\cite{mac2} to calculate the Poincare (Hilbert) series $P_n(\tau)=\sum_k \dim ^nM_k \tau ^k$ of $R$-module $^nM=\sum _k H^{k,n}$.
Here $R=\mathbb{C}[t^1,\cdots,t^{\alpha},\cdots]=\sum_m \Sym^m S.$ We calculate  generators  of this module and generators of free resolution
$$\cdots \to {^nM_i} \to \cdots \to {^nM_1}\to {^nM_0}\to {^nM}\to 0$$
where $^nM_i={^n\mu_i}\otimes R.$

Input:

Coefficients $\Gamma^m_{\alpha\beta}$ in the differential, the number of Greek indices ($\dim S$), the number of Roman indices ($\dim V$).

Output:

Poincare (Hilbert) series,\\
number of generators of $^nM$ and the number of them,\\
number of generators of $^n\mu_i$ having given degree.

\item\label{step:LiE} Using LiE, we decompose $\Sym^mS\otimes\wedge^kV$ into irreducible representation of $Aut$. Applying principle of maximal propagation and resolving the ambiguities from the information about Poincare series we obtain decomposition of cohomology into irreducible representation for $k\geq 20$.

\item\label{step:mathematica} We make a conjecture of the decomposition of $H^{k,n}$ into irreducible representation for arbitrary $k$ using the information from the step~\ref{step:LiE}. We  prove that our conjecture gives the right Poincare series using Weyl dimension formula.{\footnote{The Mathematica code for $10D$ case is provided here: \url{http://lifshitz.ucdavis.edu/~rxu/code/cohom/dim10dredux.nb}}}

\item We make a conjecture about cohomology generators using the information about their numbers and dimension from Macaulay2~\cite{mac2} and the information from the steps \ref{step:LiE} and \ref{step:mathematica}. We  prove that our formulas give cocycles using Gamma~\cite{GAMMA}.

\item We use the formula Eq.~\ref{Eq:alt_mu} to get the decomposition of generators of free resolution into irreducible representation.

\end{enumerate}


\end{document}